\documentclass[11pt,a4paper]{article}
\pdfoutput=1
\usepackage{jheppub}

\def\arXiv#1{\href{http://arxiv.org/abs/#1}{arXiv:#1}}
\def\arXiv#1#2{\href{http://arxiv.org/abs/#1}{arXiv:#1}}

\def\doi#1#2{\href{http://doi.org/#1}{#2}}

\usepackage{multirow, graphicx,amssymb,url,mathrsfs,amsmath}
\usepackage{epstopdf}
\usepackage{wrapfig,boxedminipage,setspace,subfigure,epsfig}
\usepackage{amsxtra,amstext,latexsym,dsfont,amsfonts}
\usepackage{color,eucal}
\usepackage[dvipsnames]{xcolor}
\usepackage{float}
\usepackage{slashed,comment}
\usepackage{kotex}
\usepackage{verbatim}

\usepackage[utf8]{inputenc}
\usepackage{CJKutf8}

\title{\centering \Huge More on Topological Hydrodynamic Modes}

\author[a]{Wen-Bin Pan}
\author[a,b]{and Ya-Wen Sun}

\emailAdd{panwenbin18@mails.ucas.ac.cn}
\emailAdd{yawen.sun@ucas.ac.cn}

\affiliation[a]{School of Physical Sciences, University of Chinese Academy of Sciences,\\
	Zhongguancun east road 80, Beijing 100190, China}
\affiliation[b]{Kavli Institute for Theoretical Sciences, University of Chinese Academy of Sciences,\\
	Zhongguancun east road 80, Beijing 100190, China}

\abstract{Based on previous work that topologically nontrivial gapless modes in relativistic hydrodynamics could be found by weakly breaking the energy momentum conservation, in this paper, we study the holographic system which produces the same hydrodynamic modes. In the hydrodynamic system, one possibility to obtain the energy momentum non-conservation is to couple the system to external gravitational fields, i.e. to observe the system in a special non-inertial frame. Similar to what happens in the hydrodynamic system, a non-inertial frame version of holography indeed produces the same topologically nontrivial gapless hydrodynamic modes. We also generalize the study of topological modes in relativistic hydrodynamics to the case with one extra $U(1)$ current and find that more complicated topological phase diagrams could exist when we consider more possibilities of the mass terms. We also discuss the possible underlying mechanism for this topological change in the spectrum when being observed in a non-inertial reference frame.
}

\begin{document}
\maketitle

%
\section{Introduction }\label{sec 1}

It has been theoretically discovered and experimentally confirmed during the last decade that some classical systems, like topological optical/sound systems, have nontrivial topological states \cite{lu2014topological,ozawa2019topological,zhang2018topological}.
Recent studies \cite{delplace2017topological,souslov2019topological,green2020topological} has shown that hydrodynamics may be a new candidate that possesses topological nontrivial states.
Hydrodynamics is the universal low energy effective theory that describes the near equilibrium dynamics of a system of continuum in the low frequency and small wave vector limit.
Under small perturbation away from thermal equilibrium, the system would respond
and develop both sound and diffusive modes \cite{Kovtun:2012rj}.
These modes are gapless with poles at $\omega=\mathbf{k}=0$.

In recent papers \cite{Liu:2020ksx,Liu:2020abb}, we proposed that the gapless relativistic hydrodynamic modes could become topologically nontrivial gapless modes after weakly breaking the conservation of energy momentum. Moreover, the conservation breaking and the topological modes could be observed in a non-inertial reference frame of an accelerating observer moving along a helix(or accelerating along the $x$ axis and at the same time rotating its axes when the radius of the helix is zero). 
The change in the mode structure could be considered as an effect of the inertial forces that would be observed in a non-inertial frame. The inertial forces would appear in the conservation equations as terms that are similar to interaction terms. This would modify the equations and therefore modify the eigenmodes of the hydrodynamic system. A similar effect has been found in a study on equatorial waves \cite{delplace2017topological}, which showed that the original gapless hydrodynamic modes are gapped by the Coriolis force in the rotating frame of earth and the new modes are topologically non-trivial with non-zero first Chern numbers. A different way to explain why frame transformation would change the mode structure could be found in section \ref{subsec 3.2.1} in which an infinitesimal Lorentz boost frame is considered.
There are many more aspects to explore about this interesting discovery, especially the holographic realization of the modes and the case when the system has more conserved currents (e.g. with an extra U(1) symmetry).

In this paper, we will first review the topological hydrodynamic modes and the calculation of the corresponding topological invariants originally presented in \cite{Liu:2020abb}. We further demonstrate that in order to see the non-trivial modes, an observer should also rotate themselves in a certain way along the helix. The main work in this paper will focus on the holographic realization of the topological modes and generalization of the system to the case with an extra $U(1)$ symmetry. 

To get the dispersion relations of the hydrodynamic modes in holography, we could compute the retarded Green's function of the components of the energy momentum tensor $T^{\mu \nu}$, which could be done using the GKPW rule of AdS/CFT. We add perturbations $h_{\mu \nu}$ on top of the bulk metric, and solve the Einstein equations of $h_{\mu \nu}$ with the boundary values of $h_{\mu \nu}$ being the source of $T^{\mu \nu}$.

The original holographic computation of the Green's function is adapted to an inertial frame \cite{Policastro:2002se,Policastro:2002tn}, 
while directly solving Einstein equation in a non-inertial frame is much more complicated since all the ten components of the perturbations are coupled in the equations.
However, note that to get the hydrodynamic modes, what we need is actually the pole of the correlators while not the explicit expressions of the Green's functions. Thus we find a way to circumvent solving the Einstein equation and give a prescription to obtain the poles directly without solving the equations. We transform the solutions of inertial frame $h_{\mu \nu}$ to the non-inertial reference frame and the poles of the correlators could be found from the poles of integration constants of the transformed solutions. This prescription should in principle work for any reference frame. We will first check its effectiveness in the simple Lorentz boosted frame and then apply the prescription to the helix frame (a specific helix moving frame as we will explain later), which is the reference frame that could produce topologically nontrivial gapless hydrodynamic modes. In both frames, it has been confirmed that the hydrodynamic result of the poles matches the computation following our holographic prescription. This further confirms that a non-inertial reference frame holographic system would provide a holographic dual description of the topologically nontrivial hydrodynamic system.

We also generalize the topologically nontrivial hydrodynamic system to the case with an extra $U(1)$ symmetry in this paper. In this case, more complicated topological structure and phase diagrams would arise and there are more ways to break both the energy momentum conservation and the $U(1)$ current conservation. 

This paper is organised as follows. In section \ref{sec 2}, we give a brief review of relativistic hydrodynamics and review its topological non-trivial modes which could be observed in a specific non-inertial reference frame.
In section \ref{sec 3} we introduce our prescription for the holographic computation of the poles and the explicit calculation of the poles of the Green's function both in a Lorentz boosted frame and in the helix frame, the result of which is the same as the boundary computation in section \ref{subsec 2.2}. Section \ref{sec 4} generalizes the physics in section \ref{sec 2} to the case with an extra $U(1)$ current. Section \ref{sec 5} is devoted to conclusions and discussions.

%
\section{Review of topological hydrodynamic modes under weakly breaking of conservation }\label{sec 2}

 In this section, we first briefly review some key aspects in hydrodynamics, including conservation equations, constitutive equations, hydrodynamic modes and their dispersion relations. Then in analogy with the Schrödinger equation, we introduce the notion of effective Hamiltonian in hydrodynamics for the convenience of further calculations in topological hydrodynamic modes. In section \ref{subsec 2.2}, we will review the topological hydrodynamic modes caused by a change of reference frame {and in section \ref{subsec 2.3} we will review the calculation of the topological invariants of the modes.} Besides the trajectories of the accelerating observer, we also work out the rotation of the observer's orthonormal basis as seen from the inertial frame in section \ref{subsec 2.4}.

\subsection{Review on relativistic hydrodynamics}\label{subsec 2.1}
Hydrodynamics is the universal effective description for a classical or quantum system that is close to local thermal equilibrium at long time and distance scale. In the following we will first review the basic equations in relativistic hydrodynamics, especially focusing on the spectrum of the hydrodynamic modes, or in other words, the dispersion relation of a relativistic hydrodynamic system.
In this section, we focus on relativistic hydrodynamic systems without extra internal charges. 
The system with extra conserved internal charges will be discussed in section 4. Now the only conserved quantity is the energy momentum tensor $T^{\mu \nu}$ which obeys the conservation equation
\begin{align}
\partial_{\mu}T^{\mu\nu}=0.
\end{align}
Expanding to first order in derivative expansions, the constitutive equation of $T^{\mu \nu}$ in the Landau frame is
\begin{align}
T^{\mu \nu } =\varepsilon u^{\mu}u^{\nu}+P\Delta^{\mu \nu}-\eta \Delta^{\mu \alpha }\Delta^{\nu \beta }(\partial _{\alpha }u_{\beta }+ \partial _{\beta }u_{\alpha }-\frac{2}{3}\eta _{\alpha \beta }\partial _{\lambda }u^{\lambda })-\zeta \Delta^{\mu \nu }\partial _{\lambda }u^{\lambda }+ O(\partial ^{2}) 
\end{align}
where $\Delta_{\mu \nu }=\eta_{\mu \nu }+u_{\mu}u_{\nu}$, $\varepsilon$, $P$,$\eta$ and $\zeta$ are the density, pressure, shear and bulk viscosity respectively.

Under small perturbations away from the thermal equilibrium, the system would develop hydrodynamic modes. There are four eigen-modes of the system. Two of them are the sound modes propagating in the direction of $\mathbf{k}=\left ( k_x,k_y,k_z \right ) $ with the dispersion relation $\omega= \pm v_s \left | \mathbf{k} \right | -i \Gamma_s \mathbf{k}^2$, where $\Gamma_s=(\frac{4}{3} \eta +\zeta )/(\varepsilon +P)$. 
The other two are transverse modes with $\omega=-i\frac{\eta}{\varepsilon +P}\mathbf{k}^2$.
To the first order in $\mathbf{k}$, dissipative terms disappear and the spectra of the four modes are real, which cross each other at $\omega=0$ and $\mathbf{k}=0$.

For later convenience, here in analogy with the Schrödinger equation, we develop the notion of an effective Hamiltonian in hydrodynamics following \cite{Liu:2020ksx,Liu:2020abb}.
Substituting the constitutive equations for the perturbation $\delta T^{\mu \nu}$ into the conservation equation $\partial_\mu \delta T^{\mu \nu}=0$, we could rewrite the equations by keeping the $\partial_t$ terms to the l.h.s, while putting all the other terms to the r.h.s. 
At leading order in $\mathbf{k}$, this gives us the following form of hydrodynamic equations 
\begin{align}\label{sch}
i\partial_t \Psi =H \Psi
\end{align}
where we have defined
\begin{align}
\Psi = \begin{pmatrix}
\delta \epsilon \\
\delta \pi^x\\
\delta \pi^y\\
\delta \pi^z
\end{pmatrix},
H=\begin{pmatrix}
0&  k_x&  k_x& k_x\\
v^2_s k_x&  0&  0& 0\\
v^2_s k_y&  0&  0& 0\\
v^2_s k_z&  0&  0& 0
\end{pmatrix}.
\end{align}
In this way, we have defined an effective Hamiltonian $H$ whose eigenvalues gives the spectrum of hydrodynamic modes \cite{Kovtun:2012rj}.

\subsection{Topological non-trivial modes }\label{subsec 2.2}

In the recent papers \cite{Liu:2020ksx,Liu:2020abb} we showed that it is possible to change the topologically trivial gapless modes of relativistic hydrodynamics to non-trivial ones by weakly breaking the conservation of the energy momentum tensor.
This kind of conservation breaking can be realized in a non-inertial reference frame of an observer who, in the original inertial frame, is moving along a helix. {This gives another effect for accelerating frames in addition to the Unruh effect which states that an observer uniformly accelerating in the Minkowski vacuum would observe a thermal bath of non-zero temperature.}

The main reason for the change of modes is the inertial forces that would appear in an non-inertial frame. For instance, in a rotating frame, inertial forces such as the Coriolis force and the centrifugal force would come into the conservation equations for hydrodynamics and change the mode structure. 
	
Before delving into our construction of relativistic hydrodynamics, we first give a non-relativistic example that develops topologically non-trivial modes that are similar to ours.
In \cite{delplace2017topological}, a rotating-shallow-water model was used to describe the dynamics of the equatorial waves which can be modeled as excitations of a thin layer of fluid on a two-dimensional surface of height $h(\mathbf{x},t)$ and horizontal velocity $\mathbf{u}(\mathbf{x},t)$
\begin{align}
	\partial_{t} h+\nabla \cdot(h \mathbf{u})=0
\end{align}
\begin{align}\label{coriolis}
	\partial_{t} \mathbf{u}+(\mathbf{u} \cdot \nabla) \mathbf{u}=-g \nabla h-f \hat{\mathbf{n}} \times \mathbf{u}
\end{align}
where $f=2 \Omega \cdot \hat{\mathbf{n}}$ is the Coriolis parameter, $\Omega$ is the rotation vector of the earth and $\hat{\mathbf{n}}$ is the local vertical unit vector.
Note the second term on the RHS of \eqref{coriolis} is the Coriolis force that comes from the earth's rotation and would not appear if the observer was in an inertial frame.\footnote{If an observer is rest in a frame rotating with non-zero angular velocity $\vec{\omega}$ relative to an inertial frame and the velocity of the fluid is not parrallel to the direction of the angular velocity, the Coriolis force would be observed to influence the fluid. This can be deduced from the Newton's second law of motion. By performing a transformation to the rotating frame, the acceleration $a=\frac{\mathrm{d}^2 \mathbf{x}}{\mathrm{d} t^2}$ in the inertial frame becomes $a'=a-\vec{\omega} \times (\vec{\omega} \times \mathbf{x})-2 \vec{\omega} \times v'$ in the new frame, where the Coriolis term is $-2 \vec{\omega} \times v'$ and $v'$ is the fluid velocity relative to the rotating frame. } 
After linearizing the equations about the mean height ($h=H$) and a state of rest ($\mathbf{u}=0$), we would find the equations may be rewritten in the momentum space as $i \partial_t \Psi= \mathcal{H} \Psi$ with $\Psi=(h-H,\mathbf{u})$ and
\begin{equation}
	\mathcal{H} =\begin{pmatrix}
	0&  -k_x H& -k_y H\\
	-k_x g&  0& i f\\
	-k_y g&  -i f& 0
	\end{pmatrix}.
\end{equation}
There are three eigenmodes denoted by $\Psi_{\pm}$ and $\Psi_0$ with frequencies $\omega_{\pm}=\pm\left(f^{2}+c^{2} k^{2}\right)^{1 / 2}$ and $\omega_0=0$ respectively.
Thus, the original gapless hydrodynamic modes in an inertial frame are gapped by the Coriolis force in the rotating frame of earth. 
The study further showed that each of the three modes $\Psi_{\pm}$ and $\Psi_0$ is associated with a Chern number calculated by a surface integral of Berry curvature in the parameter space $(f/c, k_x, k_y)$.
The modes $\Psi_{\pm}$ are topologically non-trivial since they have non-zero Chern numbers $\pm 2$.

In our construction for relativistic hydrodynamics, the non-inertial frame is encoded in the reference frame transformation introduced in \cite{Liu:2020ksx}
\begin{align}\label{CT0}
&x^{\mu }\to x'^{\mu}=x^{\mu }+\xi ^{\mu }\\
&\xi ^{\mu } =(-\frac{1}{2} mxt,\frac{1}{4} mx^{2} +\frac{1}{4}m v_{s}^{2}t^{2} ,-\frac{1}{2}bv_{s}zt,\frac{1}{2}bv_{s}yt),
\end{align}
where $b$ and $m$ are both infinitesimal parameters of the same order.
The coordinate transformation would induce a change of metric from the original flat metric $\eta_{\mu \nu }$ to $g_{\mu \nu }=\eta_{\mu \nu } + h_{\mu \nu }$.
Thus $T^{\mu \nu }$ would become covariant conservative in the new frame
\begin{align}\label{2.7}
\nabla'_{\mu}T^{\mu\nu}(x') & = 0
\end{align}
instead of $\partial_{\mu}T^{\mu\nu}(x)=0$. Expand the covariant conservation equation \eqref{2.7} and we get
\begin{align}\label{2.8}
\partial_{\mu} \delta T^{\mu \nu} & = -\frac{1}{2} \partial_{\alpha} h \delta T^{\alpha \nu}-\frac{1}{2} \eta^{\nu \beta}\left(2 \partial_{\mu} h_{\alpha \beta}-\partial_{\beta} h_{\mu \alpha}\right) \delta T^{\mu \alpha}.
\end{align}
The r.h.s of the above equation can be seen as the term that breaks the conservation of $T^{\mu \nu}$.

Note here the configuration that we are perturbing on is the same equilibrium state at rest in the inertial frame, which satisfies \eqref{2.7} in the non-inertial frame. Thus, the fluid we are considering stays at rest in the original inertial frame while accelerates in the non-inertial frame due to the inertial forces. This is similar to the fact that in Newton mechanics an object at rest in an inertial frame would be observed rotating in a non-inertial rotating frame, which could be viewed as the consequence of the inertial force in the non-inertial frame. To be more precise, in the inertial frame the equilibrium state is the fluid at rest with four velocity $u^\mu=(1,0,0,0)$ and at thermal equilibrium with thermodynamical variables: a uniform temperature $T$, a uniform energy density $\epsilon$ and uniform pressure density $P$. The energy momentum tensor of the background equilibrium state is
\begin{align}
	T^{\mu\nu}=\epsilon u^\mu u^\nu+P(u^\mu u^\nu+\eta^{\mu \nu})=
	\begin{pmatrix}
	\epsilon &  &  & \\
	&P &  & \\
	&  &P & \\
	&  &  &P
	\end{pmatrix},
\end{align}
which satisfies the conservation equation: $\partial_\mu T^{\mu \nu}=0$. 
In the non-inertial frame \eqref{CT0}, the equilibrium state that we perturb on is the same state but observed by the non-inertial observer. Consequently, in the new frame, the four velocity of the fluid is $u'^\mu=\frac{\partial x'^\mu}{\partial x^\alpha}u^\alpha$, the metric is $g^{\mu \nu}=\frac{\partial x'^\mu}{\partial x^\alpha}\frac{\partial x'^\nu}{\partial x^\beta}\eta^{\alpha \beta}$ and the energy momentum tensor of the background equilibrium state becomes (up to the first order in $m$ and $b$)
\begin{align}
	T^{\mu\nu}(x')=&\epsilon u'^\mu u'^\nu+P(u'^\mu u'^\nu+g^{\mu \nu})\notag\\=&
	\begin{pmatrix}
	(1-m x')\epsilon  &-\frac{1}{2}m t'(P-v_s^2\epsilon)  &-\frac{1}{2}b v z' \epsilon  &\frac{1}{2}b v y' \epsilon \\
	-\frac{1}{2}m t'(P-v_s^2\epsilon) &(1+m x')P &  & \\
	-\frac{1}{2}b v z' \epsilon &  &P & \\
	\frac{1}{2}b v y' \epsilon &  &  &P
	\end{pmatrix}
\end{align}
and from this expression we could see that the temperature, energy density and pressure density are no longer uniform in the new frame. It is this energy momentum tensor that solves the conservation equation \eqref{2.7} in the non-inertial frame.
Of course as discussed in \cite{Liu:2020ksx}, it is also possible to consider the case where the background state that is perturbed on to be the fluid at rest in the new frame with uniform temperature, energy and pressure density, which however does not obey the equation of motion thus needs an extra external force. We do not discuss this possibility here.

Then we continue to solve for perturbations of the system. By substituting the transformation \eqref{CT0} into \eqref{2.8}, we obtain
\begin{eqnarray}
\partial_{\mu} \delta T^{\mu t} & = & m \delta T^{t x} \label{t1}\\
\partial_{\mu} \delta T^{\mu x} & = & -m v_{s}^{2} \delta T^{t t} \label{t2}\\
\partial_{\mu} \delta T^{\mu y} & = & b v_{s} \delta T^{t z} \label{t3}\\
\partial_{\mu} \delta T^{\mu z} & = & -b v_{s} \delta T^{t y}. \label{t4}
\end{eqnarray}
Then the corresponding effective Hamiltonian by the definition in the last section gives
\begin{align}\label{H}
H & = \left(\begin{array}{cccc}
0 & k_{x}+i m & k_{y} & k_{z} \\
\left(k_{x}-i m\right) v_{s}^{2} & 0 & 0 & 0 \\
k_{y} v_{s}^{2} & 0 & 0 & i b v_{s} \\
k_{z} v_{s}^{2} & 0 & -i b v_{s} & 0
\end{array}\right).
\end{align}
The eigenvalue of $\omega$ satisfies
\begin{align}
\omega^2(\omega^2-k_x^2 v_s^2-k_y^2 v_s^2-k_z^2 v_s^2-m^2 v_s^2)+ v^2 b^2(k_x^2 v_s^2+m^2 v_s^2-\omega^2)=0
\end{align}
and the spectrum of $H$ is 
\begin{align}
\omega & = \pm \frac{1}{\sqrt{2}} \sqrt{b^{2}+k^{2}+m^{2} \pm \sqrt{\left(k_{x}^{2}+m^{2}-b^{2}\right)^{2}+\left(k_{y}^{2}+k_{z}^{2}\right)^{2}+2\left(k_{y}^{2}+k_{z}^{2}\right)\left(k_{x}^{2}+m^{2}+b^{2}\right)}}.
\end{align}
In the case of $k_y=k_z=0$, the eigenvalue equation and the spectrum of $H$ reduce to
\begin{align}
(\omega^2-v_s^2 b^2)(\omega^2-k_x^2 v_s^2 -m^2 v_s^2)=0\label{eigen}\\
\omega_{1,2} = \pm b v_s, \;\;\; \omega_{3,4} = \pm v_s\sqrt{k_x^2+m^2}\label{sp}.
\end{align}
This spectrum is plotted in figure. 1 for different parameters of $b$ and $m$. More details of this spectrum and the topological structure of the band crossings as well as the topological phase transitions could be found in \cite{Liu:2020ksx,Liu:2020abb}.  In order to explain why the mode \eqref{sp} is a "topologically non-trivial" spectrum, we will also review briefly the calculation of the topological invariants in the next subsection and demonstrate that it is indeed topologically non-trivial with a protecting symmetry.

\subsection{Topological invariants }\label{subsec 2.3}

{A topological invariant for the modes of a physical system is a quantity or a property that is preserved under small perturbations that do not lead to a topological phase transition. For gapless modes, the band crossing node of topological modes could not be gapped by small perturbations and is associated with nontrivial topological invariants. In comparison, the node of trivial modes is accidentally crossed and could be gapped by an arbitrarily small perturbation, thus has trivial topological invariants.

The calculation of the topological invariant of gapless systems depends on the symmetry and the dimension of the system. For the Weyl-semimetal, the topological invariant is the integration of the Berry curvature on a two dimensional sphere enclosing the Weyl point in the three dimensional momentum space, see for e.g.\cite{Liu:2018djq,Landsteiner:2019kxb}. For a Nodal line semimetal, the Berry phase is integrated on a one dimensional circle which links the circle of the nodes, see for e.g.\cite{fang2016topological,Liu:2018bye,Liu:2018djq}. For symmetry protected gapless topological states, the symmetry may reduce the effective spatial dimension of the momentum space, since we have to calculate the topological invariant on high symmetric points in momentum space. Thus, the topological invariant may need to be calculated on an even lower dimensional manifold, e.g. zero dimensional manifold.}

\begin{figure}[htbp]
    \centering
    \subfigure[]{
    \begin{minipage}[b]{.3\linewidth}
    \centering
    \includegraphics[scale=0.3]{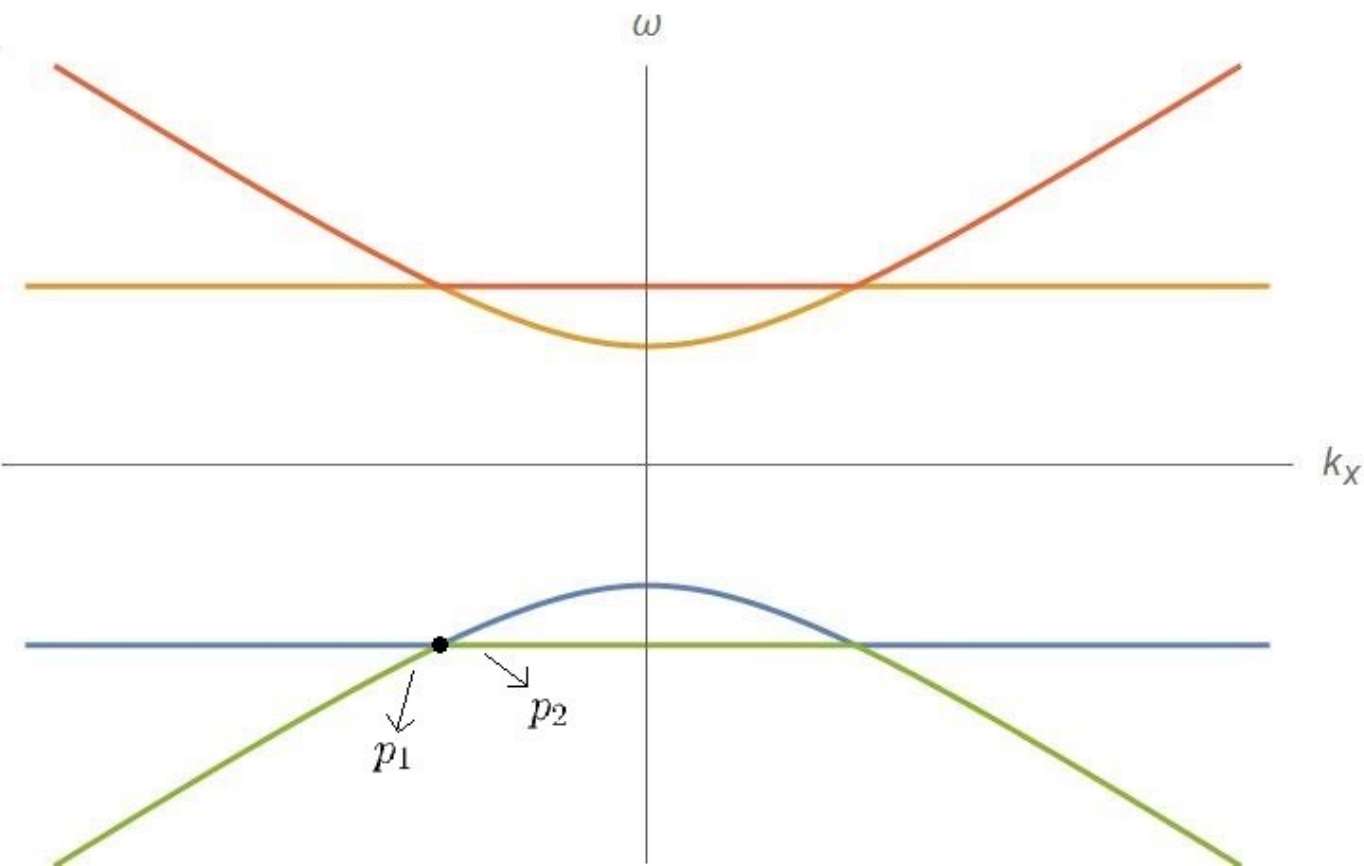}
    \end{minipage}
    }
    \subfigure[]{
    \begin{minipage}[b]{.3\linewidth}
    \centering
    \includegraphics[scale=0.3]{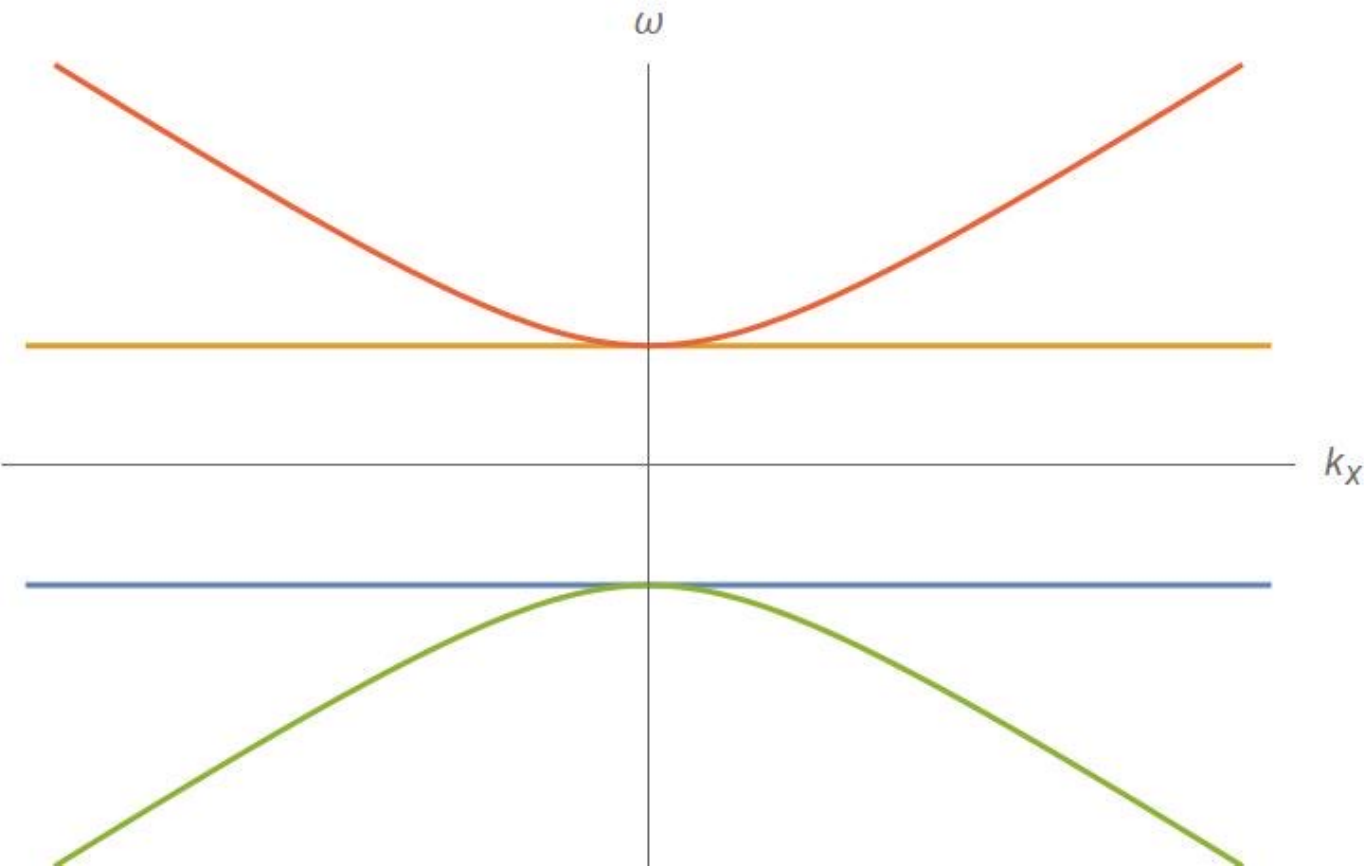}
    \end{minipage}
    }
    
    \subfigure[]{
    \begin{minipage}[b]{.3\linewidth}
    \centering
    \includegraphics[scale=0.3]{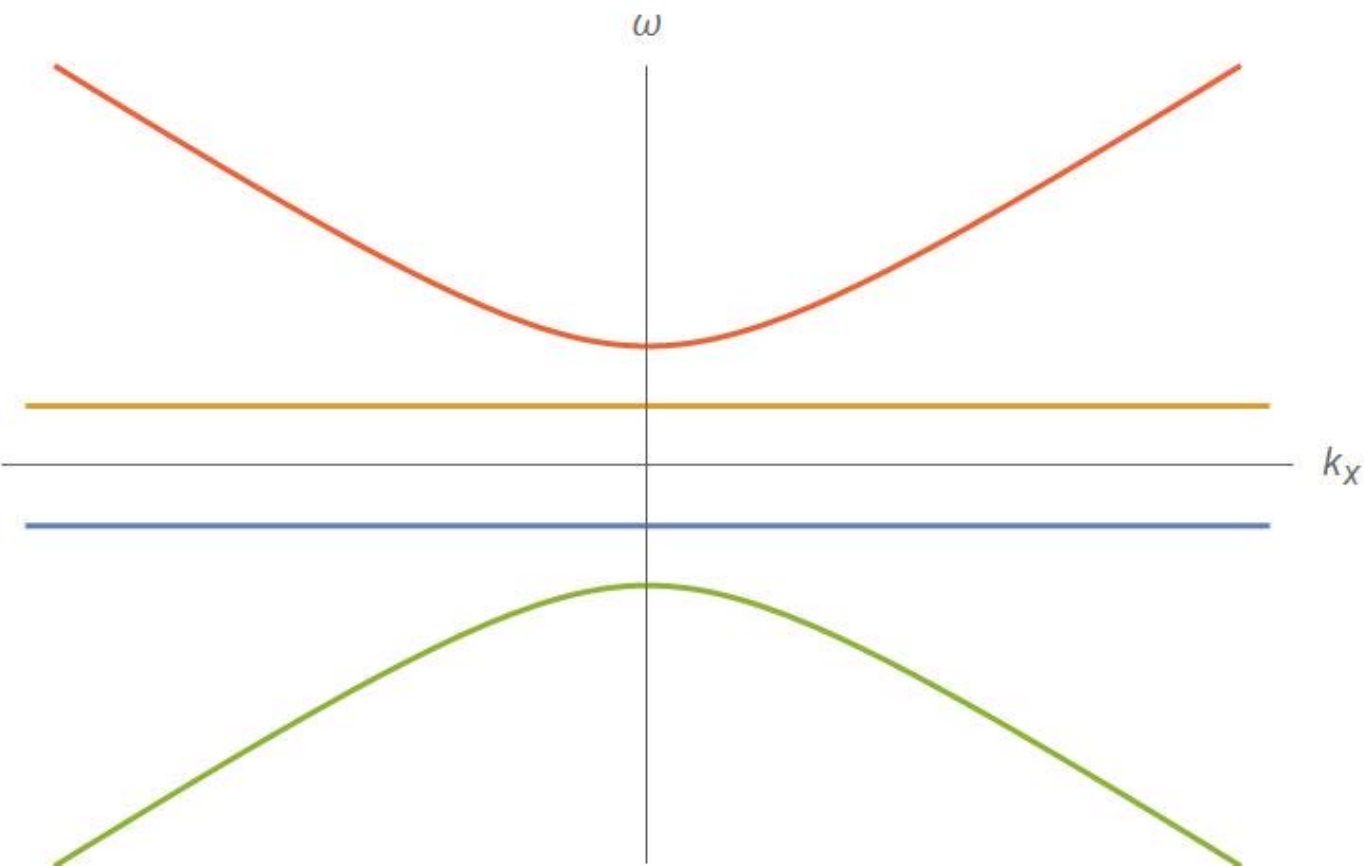}
    \end{minipage}
    }
    \subfigure[]{
    \begin{minipage}[b]{.3\linewidth}
    \centering
    \includegraphics[scale=0.3]{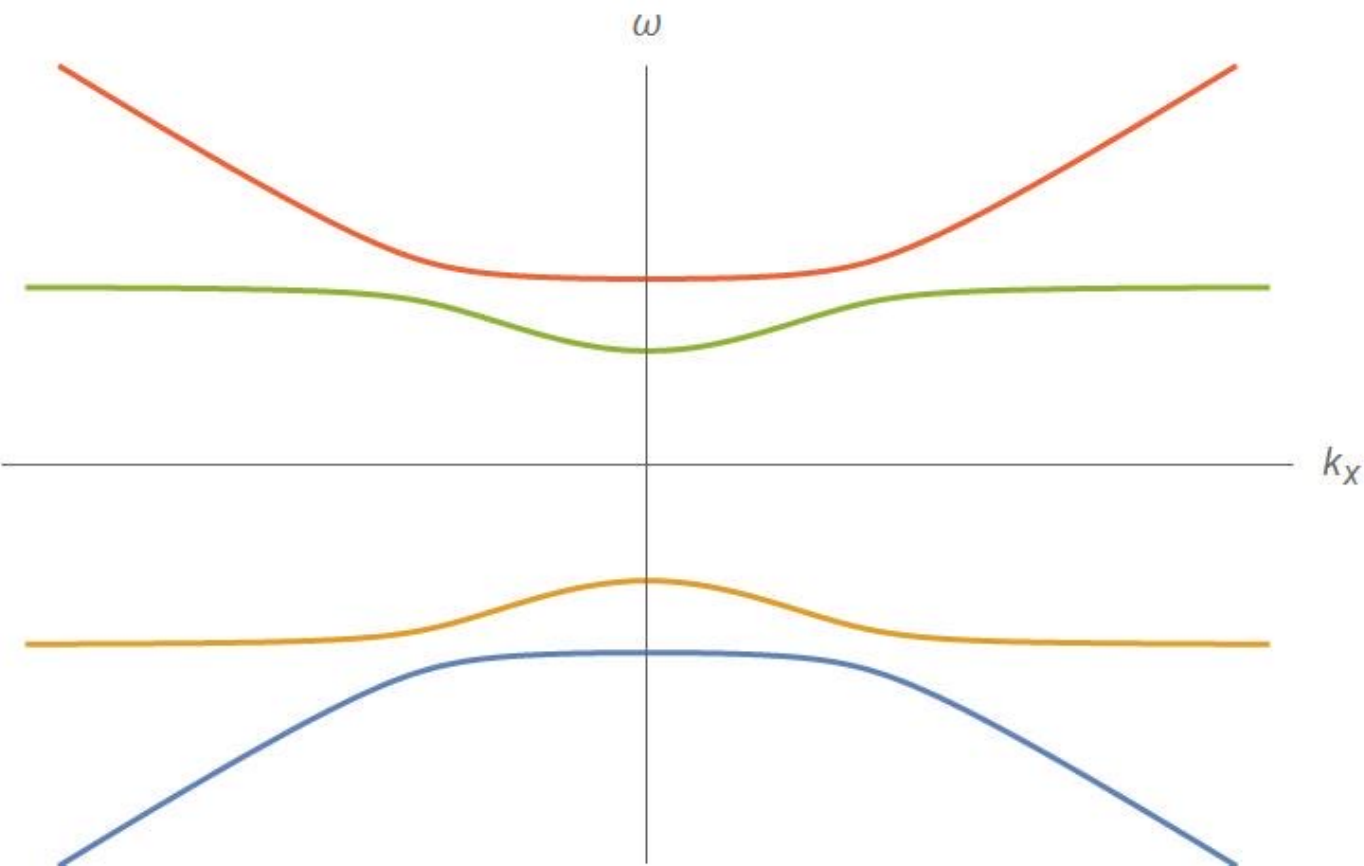}
    \end{minipage}
    }
    \caption{The topological spectrum of the Hamiltonian \eqref{H} for different relative sizes of $m$ and $b$ with $v_s=1$ and $k_y=k_z=0$. Top left: $m<b$; top right: $m=b$; bottom left: $m>b$; bottom right: an extra $m'$ term in the $y$ direction though $m<b$ (i.e. the $ty$ and $yt$ components of the Hamiltonian \eqref{H} become $k_y+im'$ and $(k_y-im')v_s^2$ respectively). At $m<b$, the modes have four band crossing nodes while at $m>b$ there are no band crossing nodes. As long as the $m'$ term in the $y$ direction is turned on, the band crossing nodes disappear. This is the consequence of the fact that the band crossing nodes in this system are topologically nontrivial protected by the reflectional symmetry in the $y$ and $z$ directions.}
    \label{spectrum}
\end{figure}

Figure \ref{spectrum} shows the dispersion relations of the hydrodynamic modes of the system.
It shows that our hydrodynamic system is gapped if $b<m$ and is gapless with four nodes at non-zero values of $k_x$ and $k_y=k_z=0$ if $b>m$. However, the nodes would disappear when $m$ terms are added in the $y$ and $z$ directions, as shown in $(d)$ of figure \ref{spectrum}.
Therefore the nodes are topologically nontrivial protected by the reflection symmetry $M$ of $y \to -y $ and $z \to -z$.
This effectively reduces the spatial dimension to be one dimension and the high symmetric points are the points on the $k_x$ axis in the momentum space. More details could be found in \cite{Liu:2020ksx,Liu:2020abb}.
Note here that the topologically protected modes are the bandcrossing nodes in this figure, which do not disappear under small perturbations of the system. In this sense the hydrodynamic system would be called trivial when there are no bandcrossings in the spectrum or when the bandcrossings are accidental touchings that would disappear under small perturbations.

{Then we could use two different but equivalent methods to compute the topological invariants of these reflection symmetry protected topological hydrodynamic modes. We could focus on the left lower node in the following and denote the left limit and right limit of the node as $p_1$ and $p_2$ respectively as shown in $(a)$ of figure \ref{spectrum}.}

The first method is explained in Sec.II A of \cite{fang2016topological}, where the topological invariant is defined as
\begin{align}
\xi = N_1-N_2,
\end{align}
in which $N_i$ is the number of occupied bands that have eigenvalue $+1$ of the reflection symmetry $M$ at point $p_i$.
In figure \ref{spectrum} the occupied state at point $p_i$ is 
\begin{align}
\left | n_1  \right \rangle &= \frac{1}{\sqrt{1/v^2_s-1}}(\frac{\sqrt{m^2+k^2_x}}{v_s(m+ik_x)},i,0,0)\\
\left | n_2  \right \rangle &= \frac{1}{\sqrt{2}}(0,0,-i,1).
\end{align}
$\left | n_1  \right \rangle $ has an eigenvalue of $+1$ under the reflection symmetry $M$ and $\left | n_2  \right \rangle$ has an eigenvalue of $-1$ under $M$. Thus the topological invariant at the left lower node is $\xi=N_1-N_2=+1$, in contrast with the trivial value of $0$.

The second method is to calculate the Berry phase of the modes.
Although the manifold we need to calculate the Berry phase is $0$ dimension, we could still define a Berry phase between the two states by 
\begin{align}
e^{-i \alpha}=\frac{ \langle n_1 | n_2  \rangle}{|\langle n_1 | n_2  \rangle|}.
\end{align}
Because $\langle n_1 | n_2 \rangle =0 $, the Berry phase corresponding to the node is undetermined.
This means that the two states cannot be connected without passing through a singular node and the two bands cannot be gapped without going through a topological phase transition.
In fact, if we keep increasing the parameter $b$, there would be a topological phase transition from the gapped phase of no band crossing ($b<m$) to a phase with four band crossing nodes ($b>m$).

This result of the two methods above confirms that the four nodes in the hydrodynamic case are topologically nontrivial protected by the reflection symmetry in the y and z directions. In this sense we call the hydrodynamic modes in the reference frame \eqref{CT0} topological nontrivial.

\subsection{The non-inertial reference frame }\label{subsec 2.4}

In \cite{Liu:2020ksx}, we worked out the non-inertial reference frame from the infinitesimal reference frame transformation \eqref{CT0}. We found that the observer who moves along some specific helix could see the spectrum \eqref{CT0}, i.e. accelerating in the $x$ direction with a constant acceleration $ a=-\frac{1}{2}mv_s^2 $ and simultaneously moving in a circle centered at the origin in the $y-z$ plane with a constant angular velocity $\omega =\frac{1}{2} b v_s$(or only accelerate in the $x$ direction at $y=z=0$ when the radius of helix is zero while rotating its aces continuously during this process).

We would also need to obtain the change of the orthonormal basis of the observer's frame along with time. The basis vectors can be naturally chosen to be the coordinate basis.
Note that in the language of differential geometry, we denote the coordinate basis as $\hat{e}_{\mu} = \frac{\partial}{\partial x^{\mu }} $.
Due to the infinitesimal coordinate transformation \eqref{CT0}, and using the chain rule of partial derivatives, the relation between the new and old coordinate basis at leading order in $b$ and $m$ is given by
\begin{align}
\frac{\partial }{\partial x'} & = (1-\frac{1}{2} m x)\frac{\partial }{\partial x}\\
\frac{\partial }{\partial y'} & = \frac{\partial }{\partial y}-\frac{1}{2} b v_s t\frac{\partial }{\partial z}\\
\frac{\partial }{\partial z'} & = \frac{1}{2} b v_s t\frac{\partial }{\partial y}+\frac{\partial }{\partial z}.
\end{align}
From the equations above, we can see that the basis vectors $\hat{e}_{y}$ and $\hat{e}_{z}$ remain orthogonal. They rotate slowly along $\hat{e}_{x}$ with an angular velocity of $b$, while the magnitude of $\hat{e}_{x}$ changes by a factor of $1-\frac{1}{2} m x$ and the direction remains the same. This means that the position at $x=0$ would be the most appropriate place for the observer. 

Thus, besides moving along a helix, to see the topological spectrum an observer should also simultaneously rotate themselves (i.e. the rotation of basis vectors $\hat{e}_{y}$ and $\hat{e}_{z}$ )with the same angular velocity as the circular movement in the $y$ and $z$ directions. 
Although all observers moving along helices of different radius could see the topological modes, the one moves along the helix of zero radius has the simplest trajectory. In the following, we will choose our observer to move in this simplest trajectory and also rotate himself with angular velocity of $\omega =\frac{1}{2} b v_s$ and refer to this frame as the helix frame for convenience.

%
\section{Holographic computation of the spectrum of the hydrodynamic modes}\label{sec 3}

Strongly coupled hydrodynamic systems could be found in holography by studying a classical gravitational theory in the bulk AdS spacetime and various interesting behavior for strongly coupled hydrodynamic systems have been studied\cite{Zaanen:2015oix,Son:2007vk,Hubeny:2011hd}. In \cite{Liu:2020abb}, a proposal for a holographic realization of the topological relativistic hydrodynamic system in section \ref{sec 2} has been proposed, where a non-inertial reference frame version of AdS/CFT correspondence was utilised and Ward identities were shown to match to each other on the two sides. In this section, we make a further step to show that the hydrodynamic modes of the two sides match to each other, too, which leads to the same spectrum as in the topological hydrodynamic system and confirms that the holographic system that we proposed is indeed dual to the topological hydrodynamic system. The holographic computation of boundary hydrodynamic modes was done in \cite{Policastro:2002se,Policastro:2002tn}, where the boundary system is the usual inertial frame one.  We need to modify their method in order to get the poles in other reference systems. We will first briefly review both their methods and results. Next we introduce our strategy for the holographic calculation of the poles in transformed reference frames starting from the example of Lorentz transformation. Then we use our strategy to compute the poles in the transformed non-inertial reference frame for which we obtained the same hydrodynamics modes as in the boundary system, though only for $k_y=k_z=0$.

\subsection{Inertial frame two point Green's function and their poles }\label{subsec 3.1}
Here we follow closely the calculation in \cite{Policastro:2002se} and \cite{Policastro:2002tn}.
According to AdS/CFT, the AdS$_5$ black hole with planar horizon is dual to a four dimensional field theory living on its boundary at finite temperature. To compute the two point Green's function of the stress-energy tensor on the boundary, we consider a small perturbation on the AdS$_5$ black hole metric
\begin{align}
g_{\mu \nu}= g_{\mu \nu}^{(0)}+h_{\mu \nu},
\end{align}
where the background AdS$_5$ black hole metric $g_{\mu \nu}^{(0)}$ is given by
\begin{align}\label{metric}
\mathrm{d}s^{2} = \frac{\pi^{2}T^{2}L^{2}}{u} (-f\mathrm{d}t ^2+\mathrm{d}x^2+\mathrm{d}y^2+\mathrm{d}z^2)+\frac{L^2}{4u^2f}\mathrm{d}u^2,   
\end{align}
and $f=1-u^2$. The perturbation $h_{\mu \nu}$ should obey the first order Einstein equations 
\begin{align}\label{Einstein}
R^{(1)}_{\mu \nu}=-\frac{4}{L^2}h_{\mu \nu}. 
\end{align}

According to the GKPW rule \cite{Gubser:1998bc,Witten:1998qj} for the perturbation of the metric
\begin{align}
\left \langle e^{\int dx h_{\mu \nu}^{0} T^{\mu \nu}} \right \rangle_{boundary} = e^{i\underline{S}_{bulk}},
\end{align}
the variation of the metric at the boundary $h_{\mu \nu}^{0}$ acts as the external source of the boundary energy–momentum tensor, thus the boundary Green's function is obtained by the second variation of the on shell action $\underline{S}_{bulk}$ with respect to $h_{\mu \nu}^{0}$.

We assume $h_{\mu \nu}$ has the form of plane waves in the four boundary directions, and without loss of generality, its propagation can be chosen towards the $z$ direction
\begin{align}
h_{\mu \nu}(t,z,u)=h_{\mu \nu}(\omega ,k_{z},u)e^{-i\omega t+ik_{3}z}.
\end{align}
$h_{u \mu}$ could be set to zero due to the gauge degrees of freedom.
Then $h_{\mu \nu}$ has ten independent components, and the equations of motion for these components can be divided into five groups that do not couple to each other. Each group of equations could be solved considered independently. There are two tensor modes: $h_{xy}$ and $h_{bb}=h_{xx}-h_{yy}$; two groups of vector modes: $h_{tx}$,$h_{xz}$ and $h_{ty}$, $h_{yz}$; and one group of scalar modes $h_{tt}$, $h_{tz}$, $h_{zz}$, $h_{aa}=h_{xx}+h_{yy}$.

We will review the computation of the Green's function of all three types of modes using the prescription for Minkowski correlators proposed in \cite{Policastro:2002se,Policastro:2002tn}  and demonstrate that the pole of the Green's function is nothing but the pole of integration constants of $h_{\mu \nu}$ that fulfills the incoming boundary condition.

\textbf{The tensor modes.} We consider the tensor mode $h_{xy}$. For simplicity, we denote $H_{xy}=h^x_y=g^{xx}h_{xy}=\frac{u}{(\pi T L)^2}h_{xy}$ and define the dimensionless energy and momentum $w=\frac{\omega}{2\pi T}$, $q=\frac{k_z}{2\pi T}$. The equation of motion for $H_{xy}$ in the momentum space is
\begin{align}
H_{xy}''-\frac{1+u^2}{u f} H_{xy}'+\frac{w^2-q^2f }{uf^2} H_{xy}=0
\end{align}
where $H_{xy}'$ is the derivative of $H_{xy}$ with respect to $u$.
The solution that satisfies the incoming boundary condition at the horizon is 
\begin{align}
H_{xy}(k,u)=C(1-u)^{-iw/2}\left [1-\frac{iw }{2} ln\frac{1+u}{2}-q^2ln\frac{1+u}{2}\right ]+O(w^2,w q^2,q^4).  
\end{align}
where $k=(w,0,0,q)$.
The term in $S_{bulk}$ that contains $H'^{2}_{xy}$ is
\begin{align}
S\propto \int dud^{4}x \frac{f}{u} H'^{2}_{xy}+ \cdots .
\end{align}
The form of the on-shell action is then 
\begin{align}
\underline{S} \propto \int d^{4}k(\frac{f}{u} H_{xy}(-k,u)H'_{xy}(k,u))\mid _{\epsilon }=\int d^{4}k H^{(0)}_{xy}(-k)(\frac{f}{u} H'_{xy}(k,u))\mid _{\epsilon },
\end{align}
where the boundary value of $H_{xy}$ is $H^{(0)}_{xy}=C(1+\frac{i w}{2}\ln2+q^2\ln2)+O(w^2,w q^2,q^4)$.
Thus in the lowest order of $w$ and $q$, we have
\begin{align}
C= H^{0}_{xy}(q).
\end{align}
The $xy-xy$ retarded Green's function is computed to the first order in $w$ and $q$ as
\begin{align}
G^{xyxy}_{R}\propto \frac{1}{H^{(0)}_{xy}} (\frac{f}{u}H'_{xy})\mid_{u=\epsilon}= iw+q^{2}.
\end{align}
We can see that for the tensor mode both the Green's function $G^{xyxy}_{R}$ and the integration constant $C$ do not have a pole.

\textbf{The vector modes.} We consider the vector modes by turning on metric perturbations $h_{tx}$ and $h_{xz}$.
Again we use $H_{tx}=\frac{u}{(\pi T L)^2}h_{tx}$ and $H_{xz}=\frac{u}{(\pi T L)^2}h_{xz}$.
The equations of motion for $H_{tx}$, $H_{xz}$ are
\begin{align}
	H_{tx}^{\prime}+\frac{q f}{w} H_{xz}^{\prime} &=0, \\
	H_{tx}^{\prime \prime}-\frac{1}{u} H_{tx}^{\prime}-\frac{w q}{u f} H_{xz}-\frac{q^{2}}{u f} H_{tx} &=0, \\
	H_{xz}^{\prime \prime}-\frac{1+u^{2}}{u f} H_{xz}^{\prime}+\frac{w^{2}}{u f^{2}} H_{xz}+\frac{w q}{u f^{2}} H_{tx}&=0 .
\end{align}
The incoming solution at the horizon is (omitting higher order in $w$ and $q^2$ terms)
\begin{align}
H^{inc}_{tx} =& -\frac{1}{2} +\frac{u^2}{2} -\frac{i}{8}(1-u^2)(4u+2\ln (\frac{1+u}{2(1-u)} ))w-\frac{u(u-2)}{4}q^2,\\
H^{inc}_{xz} =& \frac{i}{w}+\frac{1}{2}\ln(1-u^2)-\frac{i}{8}(\ln(1-u)^2+2\ln(1-u)\ln(\frac{1+u}{2})^2-8\ln(1+u)+\\
&4Li_2\frac{1-u}{2})w+\frac{1}{2}\ln(2(1+u))q^2.
\end{align}
There is also a pure gauge solution in addition to the incoming solution above. 
We denote this gauge solution of $H_{tx}$, $H_{xz}$
as $H^{I}_{tx}$, $H^{I}_{xz}$ respectively.
\begin{align}
H^{I}_{tx} = w,\\
H^{I}_{xz} = -k.
\end{align}
The full solution $H(k,u)$ is a linear combination of the two types of solutions above. Given the boundary conditions $H_{tx}(k,0)=H^{(0)}_{tx}(k)$ and $H_{xz}(k,0)=H^{(0)}_{xz}(k)$ at $u=0$, the linear coefficients can be represented by the boundary values
\begin{align}\label{C}
\left\{\begin{matrix}
H^{(0)}_{tx} = C_1 H^{inc}_{tx}\mid_{u = 0}+C_2 H^{I}_{tx}\\
H^{(0)}_{xz} = C_1 H^{inc}_{xz}\mid_{u = 0}+C_2 H^{I}_{xz}\\
\end{matrix}\right.
\Longrightarrow 
\left\{\begin{matrix}
C_1 = \frac{q^2H^{(0)}_{tx}+wqH^{(0)}_{xz}}{iw-\frac{q^2}{2} } \\
C_2 = \frac{iwH^{(0)}_{tx}+\frac{wq}{2} H^{(0)}_{xz}}{iw-\frac{q^2}{2} }.
\end{matrix}\right.
\end{align}
The term that contains $H'^{2}_{tx}$ and $H'^{2}_{xz}$ in $S_{bulk}$ is 
\begin{align}
S\propto \int dud^{4}x \frac{1}{u} (-H'^{2}_{tx}+f H'^{2}_{xz})+ \cdots 
\end{align}
The Green's functions are computed as
\begin{align}
G^{txtx}_{R}\propto \frac{1}{H^{(0)}_{tx}} (\frac{1}{u}H'_{tx})\mid_{u=\epsilon}= \frac{q^2}{iw-\frac{q^2}{2}}\\
G^{txxz}_{R}\propto \frac{1}{H^{(0)}_{xz}} (\frac{1}{u}H'_{tx})\mid_{u=\epsilon}= \frac{wq}{iw-\frac{q^2}{2}}\\
G^{xzxz}_{R}\propto \frac{1}{H^{(0)}_{xz}} (\frac{f}{u}H'_{xz})\mid_{u=\epsilon}= \frac{w^2}{iw-\frac{q^2}{2}}.
\end{align}
The poles of the Green's function are again the poles of the integration constant $C_1$ and $C_2$.

\textbf{The scalar modes.} We look into the scalar modes by turning on $H_{tt}=\frac{u}{(\pi T L)^2 f}h_{tt}$, $H_{tz}=\frac{u}{(\pi T L)^2}h_{tz}$, $H_{zz}=\frac{u}{(\pi T L)^2}h_{zz}$ and $H_{aa}=H_{xx}+H_{yy}=\frac{u}{(\pi T L)^2}(h_{xx}+h_{yy})$.
The equations 3.6, 3.7, 3.8 and 3.11 of reference \cite{Policastro:2002tn} gives the most general solutions of $H_{tt}$,$H_{tz}$,$H_{zz}$ and $H_{aa}$, which can be written as the linear combination of one incoming solution and three pure gauge solutions with coefficients $C_0$, $C_1$, $C_2$ and $C_3$.
\begin{align}
H(u)=C_0H^{inc}(u)+C_1H^{I}(u)+C_2H^{II}(u)+C_3H^{III}(u).
\end{align}
Coefficients $C_0$, $C_1$, $C_2$ and $C_3$ are determined by the boundary values $H^{(0)}_{tt}$,$H^{(0)}_{tz}$,$H^{(0)}_{zz}$ and $H^{(0)}_{aa}$.
Following a similar calculation of \eqref{C} for the vector modes, we can rewrite these coefficients using boundary values of the fields.
It turns out that they have the same denominator
\begin{align}
24(3w^2-q^2)+O(w^3,q^3).
\end{align}
This is identical to the denominator of two point Green's function given in equation 3.20 in \cite{Policastro:2002tn} up to a constant pre-factor. The computation shows again that the Green's functions in the scalar modes and the coefficients of the integration constants share the same poles.

Thus we could see that the poles of the two point Green's functions of the stress tensor could be found from the poles of the integration constants of metric perturbations determined by boundary values of the fields. This provides a quick and efficient way to get the dispersion relations of the hydrodynamic modes from the poles of the integration coefficients without fully obtaining the Green's functions.

\subsection{Computation of the poles in transformed reference frames}\label{subsec 3.2}

To calculate the hydrodynamic modes in non-inertial reference frames, we could perform a similar calculation as in section \ref{subsec 3.1} in the new background metric of the non-inertial reference frame. Here in this section, to avoid solving a lot of new equations for the perturbations in the new background, we could start with the solutions in the previous section for the inertial reference frame and transform the solutions to the new non-inertial reference frame. In this process we need to be careful to always check that the incoming boundary conditions are still satisfied. Then after obtaining the solutions in the new reference frame, we could get the poles of the Green's functions using the same method as in the previous section from the poles of the integration constants. 

The detailed procedure to calculate the hydrodynamic modes in other reference frames is as follows. In the bulk, the reference frame transformation could be generated by
\begin{align}\label{CT}
x'^{\mu }= x ^{\mu }+\xi ^{\mu }
\end{align}
where $x^\mu=(t,x,y,z,u)$, $\xi ^{\mu }$ gives the reference frame transformation and $\xi ^{u } = 0$, i.e. we have chosen to fix the radial coordinate $u$ invariant.

To find the poles of transformed two point Green's function under this change of boundary reference frame, we start from the solutions of $ h_{\mu \nu}$ in the original inertial reference frame, which can be written as a linear combination of specific solutions of Einstein equation \eqref{Einstein} and pure gauge solutions with ten coefficients $m_0,n_0,a_0,a_1,b_0,b_1,c_0,c_1,c_2,c_3$
\begin{equation}
\begin{aligned}
H_{xy} & = m_0 H_{xy}(\omega,k^3,u)\\
H_{bb} & = n_0 H_{bb}(\omega,k^3,u)\\
H_{tx} & = a_0 H_{tx}(\omega,k^3,u)+a_1 H_{tx}^I(\omega,k^3,u)\\
H_{xz} & = a_0 H_{xz}(\omega,k^3,u)+a_1 H_{xz}^I(\omega,k^3,u)\\
H_{ty} & = b_0 H_{ty}(\omega,k^3,u)+b_1 H_{ty}^I(\omega,k^3,u)\\
H_{yz} & = b_0 H_{yz}(\omega,k^3,u)+b_1 H_{yz}^I(\omega,k^3,u)\\
H_{tt} & = c_0 H_{tt}(\omega,k^3,u)+c_1 H_{tt}^I(\omega,k^3,u)+c_2 H_{tt}^{II}(\omega,k^3,u)+c_3 H_{tt}^{III}(\omega,k^3,u)\\
H_{tz} & = c_0 H_{tz}(\omega,k^3,u)+c_1 H_{tz}^I(\omega,k^3,u)+c_2 H_{tz}^{II}(\omega,k^3,u)+c_3 H_{tz}^{III}(\omega,k^3,u)\\
H_{zz} & = c_0 H_{zz}(\omega,k^3,u)+c_1 H_{zz}^I(\omega,k^3,u)+c_2 H_{zz}^{II}(\omega,k^3,u)+c_3 H_{zz}^{III}(\omega,k^3,u)\\
H_{aa} & = c_0 H_{aa}(\omega,k^3,u)+c_1 H_{aa}^I(\omega,k^3,u)+c_2 H_{aa}^{II}(\omega,k^3,u)+c_3 H_{aa}^{III}(\omega,k^3,u)
\end{aligned}
\end{equation}
where $ H_{\mu \nu} = h^{\mu}_{\nu}$, $H_{aa}=H_{xx}+H_{yy}$ and $H_{bb}=H_{xx}-H_{yy}$. We will need to first rotate the system to have the momentum $k$ in an arbitrary direction with all $k_x$, $k_y$ and $k_z$ components. Then as we have demonstrated earlier, finding the poles of the Green's functions amounts to calculating the poles of the linear coefficients determined from the boundary values of the fields. To find the exact expressions for the coefficients, we first transform the original solutions of $ h_{\mu \nu} $ to the new reference frame instead of computing the solutions of metric perturbations directly in the new reference frame. Note that the solutions above are written in momentum space but The transformation takes place in the coordinate space, so we have to figure out the corresponding momentum space transformation of $ h_{\mu \nu} $.

Once we obtain the new solutions of $ h_{\mu \nu} $ or equivalently $H_{\mu \nu}$ in the new reference frame in the momentum space still expressed using the ten coefficients above, we set $H_{\mu \nu}$'s boundary value to be $ H_{\mu \nu}(0) $. This generates ten equations for the ten independent components of $H_{\mu \nu}$ at $u=0$ in the form of
\begin{align}
\vec{H}_0 = M \cdot \vec{C} 
\end{align}
where $ \vec{H}_0 = (H_{xy},H_{xx},H_{tx},H_{xz},H_{ty},H_{yz},H_{tt},H_{tz},H_{zz},H_{yy})^T$ at the boundary $u=0$ , $ \vec{C} = (m_0,n_0,a_0,a_1,b_0,b_1,c_0,c_1,c_2,c_3)^T$.
Then we have
\begin{align}\label{pole}
\vec{C} = M^{-1} \cdot \vec{H_0}
\end{align}
and the poles of every component of $\vec{C}$ are determined by the equation $det(M)=0$. Thus the poles of the Green's function are just the roots of this equation. 

We will follow this procedure to calculate the hydrodynamic modes in the specific new reference frame proposed in section \ref{sec 2}. Before that, we will first use this procedure to calculate the spectrum for the Lorentz boost transformation and show that the result is the same as that calculated directly from a direct transformation from the spectrum.

\subsubsection{Infinitesimal Lorentz boost frame}\label{subsec 3.2.1}
The method to compute hydrodynamic modes in holography above should be applicable to any reference frame transformations and the results should be identical to the field theory computation. In order to show how the method works and also to test its correctness, let us start with a simple example: the infinitesimal Lorentz boost.  
In this example we consider the change of ideal (non-dissipative) hydrodynamic modes under a small Lorentz boost $\Lambda$ in the $z$ direction, where 
\begin{align}\label{LT}
\Lambda =
\begin{pmatrix}
\gamma &  0&  0&  -v \gamma\\
0&  1&  0&  0\\
0&  0&  1&  0\\
-v \gamma&  0&  0&  \gamma\\
\end{pmatrix}
\end{align}
The speed of the boost $v$ is an infinitesimal parameter and $\gamma=\sqrt{1-v^2}$.

Under this boost transformation, the velocity of the fluid $u^\mu$ observed in the new frame would transform from the original one according to \ref{LT} and there will be a global energy flux along the $z$ direction.
Different from the non-inertial reference frame case in the next subsection, here without loss of generality, we could still choose the wave vector of the mode to be in the $z$ direction, i.e. $\vec{k} =\left ( 0,0,k \right )$ and we do not need to first rotate the system to transform the momentum to a general direction.

Before the boost transformation, since there is no viscosity in ideal hydrodynamics, the dispersion relation of the mode is 
\begin{eqnarray}
\omega & = & \pm v_s k\\
\omega & = & 0.
\end{eqnarray} 
The dispersion relation shows the poles of the Green's functions for the components of energy–momentum tensor.
This means the denominator of the Green's functions is proportional to
\begin{align}
\omega^2(\omega^2-v_s^2 k^2).
\end{align}

The boost transformation changes the system to a new reference frame which moves at a constant velocity compared to the original frame. As both reference frames are inertial frames, the equations of motion should not change so on-shell modes would still be on-shell modes. However, the exact expression for the spectrum would still get modified being transformed to the new inertial frame because the values of $w$ and $k$ have changed when being observed by new observers due to the velocity difference. Another way to see this point is to directly calculate the spectrum in the new inertial frame with the values of $w'$ and $k'$ of the new frame. One would naively thinks that the spectrum should not change because the equations of the hydrodynamics should stay the same under a Lorentz transformation, however, in fact, the velocity of the fluid changes during this process, i.e. the fluid at rest in the original inertial frame now has a collective velocity in the new inertial reference frame. Thus when we calculate this spectrum directly in the new inertial frame, the result would also get modified. 

In the following we will first calculate the spectrum in the Lorentz transformed inertial frame using the two ways above and show that they give identical results. Then we show that we could get the same modified spectrum in the new inertial frame in holography using the procedure of \eqref{CT} to \eqref{pole} at the beginning of section \ref{subsec 3.2} as a consistency check of our holographic calculation. 

\vspace{0.2cm}
\noindent{\bf  I. Spectrum obtained from transforming $w$ and $k$.}

As on-shell modes are still on-shell under inertial frame transformations, we could get the spectrum in the new frame by transforming the spectrum in the original frame to the new one. Under the Lorentz boost $\Lambda$, the frequency $\omega$ and the wave vector $\vec{k}=(0,0,k)$ transform as $\omega'= \gamma \omega- v \gamma k$ and $k'= -v \gamma \omega + \gamma k$ respectively. 
Therefore the denominator of the Green's functions should change by replacing $\omega$ and $k$ in the original expression by $\omega'$ and $k'$ and it becomes
\begin{align}\label{dem}
\frac{(\omega'+k' v)^2 [k'^2 (v^2 - v_s^2) + 
	2 k' v (1 - v_s^2) \omega' + (1 - v^2 v_s^2) \omega'^2]}{(1 - v^2)^2}.
\end{align}
The new spectrum is therefore 
\begin{eqnarray}
\omega' & = & \frac{v_s-v}{1-v v_s} k'\label{v'}\\
\omega' & = & -\frac{v_s+v}{1+v v_s} k'\label{v''}\\
\omega' & = & - v k'\label{v}.
\end{eqnarray} 
 One could easily read out the velocity of the modes in the new frame from the coefficients of $k'$ in the three equations above and the velocities are identical to those calculated from the velocity superposition formula in special relativity.
\eqref{v'} (and \eqref{v''}) corresponds to the sound mode that moves in the opposite (and in the same) direction with the boost, while \eqref{v} corresponds to the two transverse modes that do not propagate in the original frame but propagate with the boost velocity in the boosted frame.

Since here we are working on an infinitesimal Lorentz boost, the form of which is convenient for the comparison with the holographic computation later, we expand \ref{dem} on the small boost velocity $v$ and keep only the leading order term in $v$ 
\begin{align}\label{CT1}
\omega'^2(\omega'^2-v_s^2 k'^2)+\omega' k'[-2 v_s^2 k'^2 + (4 - 2 v_s^2) \omega'^2]v.
\end{align}
This is the final expression for the denominator in the case of an infinitesimal Lorentz boost, which will be obtained again from the holographic calculation as will show later.

\vspace{0.2cm}
\noindent
{\bf II. Spectrum calculated from the new frame with transformed $u^\mu$.}

In the new boosted frame, the energy momentum tensor in the equilibrium is 
\begin{align}
T^{\mu\nu}=(\epsilon+P)u'^\mu u'^\nu+P\eta^{\mu\nu},
\end{align}
where $u'^\mu=\lambda^\mu_\nu u^\nu=(\gamma,0,0,-v \gamma)$ is the fluid velocity in the new frame,  $\epsilon$ and $P$ are the energy density and pressure in the equilibrium respectively.
We consider a small fluctuation of the system: $\epsilon \to \epsilon+\delta\epsilon$, $P \to P+\delta P$, $u'^\mu \to u'^\mu + \delta u'^\mu$, and the variance of $T^{\mu\nu}$ is
\begin{align}
\delta T^{\mu\nu}=(\delta\epsilon+\delta P)u'^\mu u'^\nu+(\epsilon+P)(\delta u'^\mu u'^\nu+u'^\mu \delta u'^\nu)+\delta P\eta^{\mu\nu}.
\end{align}
Note that there is a normalization constraint on the four-velocity $u'^\mu$, i.e. $u'^\mu u'_\mu = -1$.
So we have $\delta u'^t = -v \delta u'^z$.
The sound speed is the propagation speed of the fluctuation. In order to calculate the spectrum and obtain the speed of sound for this system in the new frame, we have to solve the conservation equation
\begin{align}
\partial_\mu \delta T^{\mu\nu}=0.
\end{align}
By assuming the fluctuations propagate only in the $z$ direction, we can write out the conservation equation in the momentum space explicitly as
\begin{equation}
\begin{aligned}
\partial_t [\gamma^2(1+v^2 v_s^2)\delta \epsilon-2v \gamma \pi^z]+ik^z[-v\gamma^2(1+v_s^2)\delta \epsilon+\gamma(1+v^2) \pi^z]=0\\
\partial_t \pi^x-ik^z v \pi^x=0\\
\partial_t \pi^y-ik^z v \pi^y=0\\
\partial_t [-v\gamma^2(1+v_s^2)\delta \epsilon+\gamma(1+v^2) \pi^z]+ik^z[\gamma^2(v^2+v_s^2)\delta \epsilon-2v \gamma \pi^z]=0,\end{aligned}
\end{equation}
where we have chosen the hydrodynamic variables to be $\delta\epsilon$, $\pi^i = (\epsilon+P)\delta u'^i$. $v_s^2=\frac{\delta P}{\delta \epsilon}$ is the squared sound speed in the rest frame.
{By rearranging these equations, we could obtain a form that is the same as \eqref{sch}, i.e. an effective ``Schrödinger" equation
\begin{align}
i \partial_t \begin{pmatrix}
\delta \epsilon \\
\delta \pi^x\\
\delta \pi^y\\
\delta \pi^z
\end{pmatrix}
=k^z \begin{pmatrix}
-\frac{v-v v_s^2}{1-v^2 v_s^2}&  0&  0& \frac{\sqrt{1-v^2}}{1-v^2 v_s^2}\\
0&  -v&  0& 0\\
0&  0&  -v& 0\\
\frac{(1-v^2)^{3/2}v_s^2}{1-v^2 v_s^2}&  0&  0& -\frac{v-v v_s^2}{1-v^2 v_s^2}
\end{pmatrix}
\begin{pmatrix}
\delta \epsilon \\
\delta \pi^x\\
\delta \pi^y\\
\delta \pi^z
\end{pmatrix}.
\end{align}
From this equation we could get the effective Hamiltonian $H$ defined in the section \ref{subsec 2.1}. The four eigenvalues of $H$ are $v'_s k^z$, $v''_s k^z$, $-v k^z$ and $-v k^z$.
where $v'_s=\frac{v_s-v}{1-v v_s}$, $v''_s=-\frac{v_s+v}{1+v v_s}$.
Thus we obtain the spectrum
\begin{eqnarray}
\omega & = & \frac{v_s-v}{1-v v_s} k^z\\
\omega & = & -\frac{v_s+v}{1+v v_s} k^z\\
\omega & = & - v k^z,
\end{eqnarray} 
which as expected is the same spectrum as \eqref{v'}, \eqref{v''} and \eqref{v} in the last subsection.}

\vspace{0.2cm}
\noindent{\bf III. Spectrum obtained from the holographic procedure.}

Next we check this new form of spectrum from the Lorentz boost could be reproduced from the holographic computation using the procedure at the beginning of this section. One should first check whether the solutions of metric perturbations still satisfy the incoming condition at the horizon after the Lorentz boost \eqref{LT}.
The general form of an incoming solution in the background metric \eqref{metric} can be written as 
\begin{align}
\phi(t,x,u)&\sim f(k,u)(1-u)^{-\frac{i}{2}\omega}e^{-i\omega t+ikx}\\
&=f(k,u)e^{-i\omega (t+\frac{1}{2}\ln(1-u)) +ikx}
\end{align} at the horizon.
Note that if the coefficients in front of $t$ and $\ln(1-u)$ have the same sign, $\phi $ would be incoming. 
After the transformation (\ref{LT}), $\phi $ becomes
\begin{align}
\phi'(t',x',u)&\sim f(k,u)(1-u)^{-\frac{i}{2}\omega}e^{-i\omega \gamma(t'+v x') +ik\gamma (x'+v t')}\\
              &\sim f(k,u)e^{-i\gamma (\omega-v k)t'+\frac{1}{2\gamma}\omega\ln(1-u)},
\end{align}
where in the second line the $x'$ dependence has been omitted. Here as the vector $(\omega, k)$ is time-like with $ \omega> v k$, $\omega>0$ and $\gamma>0$, it could be easily checked that $\phi'$ also satisfies the incoming boundary condition.

With the infinitesimal Lorentz boost transformation
\begin{align}
\xi ^{\mu } = (-vz,0,0,-vt,0),
\end{align} we could get the transformed metric perturbation $h_{\mu \nu } $ according to 
\begin{align}\label{h}
h'_{\mu \nu } =h_{\mu \nu } -h_{\alpha \nu } \partial_\mu \xi ^{\alpha }- h_{\mu\beta } \partial_\nu \xi ^{\beta }-\xi ^{\lambda }\partial_\lambda h_{\mu \nu }.   
\end{align} 
After we get the new expression of $h_{\mu \nu } (t,z,u)$ from the transformation above, we change it to them momentum space by Fourier transformation.
Under the Fourier transformation, terms in the form of $x^j h_{\mu \nu }$ and $\partial_{x_j}h_{\mu \nu }$ transform as 
\begin{align}
t\to -i\partial_\omega ,x_j\to i\partial_{k_j}, \partial t\to-i\omega ,\partial_{x_j}\to ik_j.
\end{align} 
Using these rules, we obtain the transformed $h_{\mu\nu}(\omega,k,u)$ fields in the momentum space. Then by setting $u=0$, the boundary values of $h_{\mu\nu}(\omega,k,0)$ are obtained and we calculate the poles of the integration constants from \eqref{CT} to \eqref{pole}. 
This amounts to calculating the determinant of the matrix $M$ and we have
\begin{align}
\det M = \frac{(kv+\omega)^2(k^2(3v^2-1)+4k\omega v+(3-v^2)\omega^2}{(1-v^2)^2}.
\end{align}
Note that $v$ is a small parameter of first order and we only keep the first order terms in our computations starting from \eqref{h}.
To the first order of $v$, it becomes
\begin{align}\label{CT12}
\omega^2(k^2-3\omega^2)+ 2\omega k(k^2 - 5\omega^2)v.
\end{align}
The speed of sound $v_s=\frac{1}{\sqrt{3}}$ in holography because of the conformal invariance of the boundary theory. Note that $\omega$ and $k$ in \eqref{CT12} and \eqref{CT1} are the frequency and wave vector observed in the boosted new reference frame.
Therefore, equation \eqref{CT12} is identical to equation \eqref{CT1}  up to a constant coefficient, which demonstrates the effectiveness of our holographic procedure for calculating the hydrodynamic modes in other reference frames.

\subsubsection{The helix frame }\label{subsec 3.2.2}
The reference frame transformation required for the nontrivial topological modes in section \ref{subsec 2.2} is generated by $x'^{\mu}=x^\mu +\xi^{\mu } $ and 
\begin{align}\label{CT2}
 \xi^{\mu } =(-\frac{1}{2} mxt,\frac{1}{4} mx^{2}+\frac{1}{4} mv_{s}^{2} t^{2},-\frac{1}{2}bv_{s}zt, \frac{1}{2}bv_{s}yt,0),
\end{align}
where $m$, $b$ are infinitesimal parameters of the same order.
We have pointed out in section \ref{subsec 2.4} that this transformation is a reference frame transformation.

Parallel to the case of Lorentz transformation, here we should also have three ways to get the new spectrum: transforming the inertial frame spectrum to the non-inertial frame, calculating the spectrum directly in the new frame, and the holographic calculation. In section \ref{subsec 3.2.1}, all these three ways have been used to calculate the spectrum for the Lorentz transformation case.
For this non-inertial frame case, the calculation of the spectrum in section \ref{sec 2} is exactly the second way above. In the following we will calculate the spectrum using the third way, i.e. holography. Before that we have some comments regarding the first way to calculate the spectrum, i.e. by transforming it from the original inertial frame. 

As we know, compared to the equations of motion in inertial frames, in non-inertial reference frames, the equations of motion will not stay invariant anymore but will get modified with extra inertial forces in the equations. Thus in the new frame the equations of motion have changed together with the constitutive equations of the background fluid. Nevertheless, we should still be able to get the modes and the spectrum of the new frame by transforming them from the original frame, i.e. substituting the transformation of $w$ and $k$ into the inertial frame spectrum to get the non-inertial frame one. However, in this case, the transformation of $w$ and $k$ is not easy to get in comparison to the Lorentz case. We could take a look at the difficulty of the transformation of $w$ and $k$ from the Fourier transformation of a plane wave as follows.

For simplicity, we set $b=0$ and keep only the $m$ terms of the reference frame transformation \eqref{CT2} and consider a plane wave in the new frame of the form
\begin{align}
f'(t',x')=e^{-i\omega't'+ik'x'}.
\end{align} 
Then in the original inertial frame, by substituting the transformation of $x'$ and $t'$, this plane wave becomes
\begin{align}\label{wave}
f(t,x)=e^{-i\omega'(t+\frac{1}{2}mxt)+ik'(x-\frac{1}{4}mx^2-\frac{1}{4}mv_st^2)}.
\end{align} 

This expression above for a plane wave of the new non-inertial frame is obviously not a plane wave in the inertial frame, i.e. in the $x$, $t$ frame. If we perform a direct Fourier transformation, the result should not be a single Fourier mode. Thus it seems that a single plane wave in the non-inertial frame could not be a single plane wave in the original inertial frame, but a combination of several modes. However, note that to get the spectrum in the second way, we have imposed a constraint in the parameter, i.e. we are working in the limit of small $m$ and $b$, thus we could still Fourier transform this function in the inertial frame to see if this is a single plane wave in this limit, i.e. if the result is a multiplication of $\delta$ functions of $\omega$ and $k$ after the Fourier transformation in the special limit. 

Again we first show this transformation in the Lorentz boost case, {in which \eqref{wave} would become 
\begin{align}
f(t,x)=e^{-i\omega'(\gamma t-v \gamma x)+ik'(-v \gamma t+\gamma x)},
\end{align}
and performing a Fourier transformation in the inertial frame, we obtain
\begin{align}
\tilde{f}(\omega,k)= 4\pi^2 \delta(\gamma k'+v \gamma \omega'-k)\delta(v \gamma k'+ \gamma \omega'-\omega).
\end{align}
Thus the procedure above gives us the correct transformation from $w$, $k$ to $w'$, $k'$ as the boost transformation is linear.}

However, for nonlinear reference frame transformations, it is not straightforward to obtain the transformed $\omega$ and $k$. 
Performing a Fourier transformation for the plane wave \eqref{wave} in the inertial frame 
\begin{align}
\tilde{f}(\omega,k)= \int \mathrm{d}t \mathrm{d}x f(t,x) e^{-(-i\omega t+ikx)},
\end{align} and we get
\begin{align}
\tilde{f}(\omega,k)= \int \mathrm{d}t \mathrm{d}x e^{-i\omega'(t+\frac{1}{2}mxt)+i\omega t+ik'(x-\frac{1}{4}mx^2-\frac{1}{4}mv_st^2)-ikx},
\end{align}
which becomes
\begin{align}
\tilde{f}(\omega,k)=\frac{4\pi }{\sqrt{m^2(w'^2-v_s^2 k'^2)}}e^{-i\frac{k^2 k' v_s^2+k'(k'^2 v_s^2+\omega^2-4\omega \omega'+3\omega'^2)-2k(k'^2 v_s^2+\omega'(\omega'-\omega))}{m(v_s^2 k'^2-w'^2)}}.
\end{align}
This cannot be identified with delta functions. An immediate reason would be that in the process to obtain the spectrum in section 2, we have been working in the limit of $b t\ll 1$ and $m x \ll 1$, $m t \ll 1$. Thus for the Fourier transformation to give the correct result, we would also need to impose such a limit in the transformation, which however, makes it difficult to get the result for the Fourier transformation. 
Without knowing this result, it is difficult to figure out the underlying mechanism for this topological change in the spectrum, however, we have the following two speculations on what is happening here.

\begin{itemize}
    \item I. It is possible that after taking into account the special limit, the results are still not $\delta$ functions, indicating that a plane wave in a  non-inertial frame is not a plane wave with a single $\omega$ and $k$ in the inertial frame, but a combination of many plane waves of the inertial frame, and vice versa. In this case, the new spectrum could only be calculated using the second and third ways and the topological change in the structure of the spectrum could only be thought of as the result of inertial forces in the non-inertial frame. 
\item II. It is possible that after taking into account the special limit, the result becomes a multiplication of $\delta$ functions. However, hints from simple quadratic terms in $x$ or $t$ in the exponent of the function tell us that now the new $\delta$ functions may imply imaginary parts in the new $\omega$ and $k$. We could see this from the following simple example. Consider a simple $t'= t- a t^2$ transformation, where $a$ is a small dimensionful parameter analogous to our parameters $b$ and $m$. Then a plane wave in the $t'$ frame becomes 
\begin{equation}
e^{-i w' t'}=e^{-i w't- i a w' t^2}=e^{-i w't- i a' t^2}
\end{equation} in the $t$ frame, where we have redefined $a'=a w'$ for simplicity. The $t$ frame Fourier transformation for the function above is
\begin{equation}
\frac{e^{-i \frac{(w'-w)^2}{4 a'}}}{\sqrt{-2i a'}}=\frac{e^{-i \frac{(w'-w)^2}{4 a'}+\frac{\pi i}{4}}}{\sqrt{ 2 a'}}.
\end{equation} The extra $\pi i/4$ factor is not important here. 
For Fourier transformations to work, we have assumed that all $w', w, k $ and $k'$ should be real, however, if we literally allow $w'$ and $w$ to be complex and write $w'=(1-i )\tilde{w'}$ and $w=(1-i)\tilde{w}$, the exponent term above would look like
\begin{equation}
\frac{e^{- \frac{(\tilde{w'}-\tilde{w})^2}{2 a'}}}{\sqrt{ 2 a'}},
\end{equation} which as we are considering the small $a'$ limit would correspond the limiting form of the $\delta(\tilde{w'}-\tilde{w})$ function. This gives some hint that it might be that this quadratic coordinate change would result in imaginary parts in the $w$ and $k$ after the transformation, however, this simple calculation is far from getting the exact imaginary parts in $w$ and $k$ that we need, 
because it only captures the zeroth order in $a'$ effect while we also require the first order in $a'$ effect, which we do not know how to get at this stage. Our speculation is that after considering the first order in $a'$ effect, when $w'$ is complex, the resulting $w$ could still be real, i.e. a real frequency plane wave in the inertial frame is still a single plane wave in the non-inertial frame but both $\omega'$ and $k'$ become complex, and vice versa. This is consistent with our computation in the case of hydrodynamics in which if we do not take the special limit, the modes of different $\omega$ and $k$ would be mixed, while when we take that limit, the modes would not be mixed and could be solved from the differential equation \eqref{sch} with the Hamiltionian \eqref{H}.

A second piece of indirect evidence comes from an example in classical mechanics. If we start from a classical damped oscillator with the Lagrangian
{\begin{equation}
L = e^{\gamma t}(\frac{1}{2}m\dot{x}^2-\frac{1}{2}kx^2),
\end{equation}
whose equation of motion is
\begin{equation}
\ddot{x}+\gamma \dot{x}+\frac{k}{m}x=0,
\end{equation}
where the dot on the top of $x$ denotes the time derivative of $x$, and $\gamma$ represents the friction of the system.
In the case of $\gamma^2<4\frac{k^2}{m^2}$, the solution of this system is  
\begin{equation}
x = C_1 e^{\frac{i}{2}(\alpha + i\gamma)t}+C_2 e^{\frac{i}{2}(-\alpha + i\gamma)t}
\end{equation}
where we have denoted $\alpha = \sqrt{\left |\gamma^2-4\frac{k^2}{m^2}\right | }$.
The frequency $\omega = \frac{1}{2}(\pm \alpha + i\gamma)$ of the solution has an imaginary part, indicating the dissipating effect.
Then we could perform a coordinate transformation (a reference frame change) of $x'= e^{\gamma t/2}x$ and find that the equation of motion for $x'$ is
\begin{equation}
\ddot{x'}+(\frac{k}{m}-\frac{\gamma^2}{4})x'=0
\end{equation}
The solution for this equation is simply the plane wave of the form
\begin{equation}
x' = C_1 e^{\frac{i}{2}\alpha t}+C_2 e^{-\frac{i}{2}\alpha t}
\end{equation}
where the frequency of the solution is real, i.e. $\omega=\pm \frac{1}{2}\alpha$. 
The imaginary part of $\omega $ disappears and there is no dissipation after the transformation. Note that at small $\gamma t\ll 1$, the reference frame transformation $x'= e^{\gamma t/2}x$ becomes $x'=x+\frac{\gamma}{2} t x$, which is of the same type of the frame transformation as our helix frame, with a $tx$ product term in the transformation. This gives us the second hint to our speculation that a reference frame transformation may turn the frequency $\omega$ from complex to real and vice versa.}

{A third piece of evidence comes from a direct computation of the spectrum of a two dimensional scalar field $\phi$ in a specific non-inertial frame associated with a uniformly accelerating observer. The transformation between this non-inertial frame and an inertial one is
\begin{equation}\label{moller}
\begin{aligned}
T &= t + a x t\\
X &= x + \frac{1}{2} a t^2
\end{aligned}
\end{equation}
where $T$ and $X$ are original inertial coordinates, while $t$ and $x$ are coordinates in the new frame.
In fact, this coordinate transformation is the infinitesimal form of the M\o ller transformation\cite{moller1952theory} which transform the inertial reference frame to a uniformly accelerating frame
\begin{equation}
\begin{aligned}
T &= (x + \frac{1}{a})\sinh(at)\\
X &= (x + \frac{1}{a})\cosh(at)-\frac{1}{a}
\end{aligned}
\end{equation}
where $a$ is the magnitude of the acceleration in the non-inertial frame.
By taking the limit of $at \ll 1$, one would find that the M\o ller transformation comes back to \eqref{moller}.
The inertial frame scalar field $\phi$ obeys the Klein-Gordon equation 
\begin{equation}
\partial_\mu \partial^\mu \phi - m^2 \phi=0,
\end{equation}
while in the new accelerating frame, this equation would become
\begin{equation}\label{scalar eom}
\begin{aligned}
\nabla_\mu \nabla^\mu \phi - m^2 \phi &= \partial_\mu \partial^\mu \phi + \Gamma^{\mu}_{\mu \lambda}(\partial^\lambda \phi)- m^2 \phi\\ &= \partial_\mu \partial^\mu \phi + \frac{1}{2} \eta^{\mu \rho} (\partial_\lambda h_{\mu \rho})(\partial^\lambda \phi)- m^2 \phi \\ &= (-\partial^2_t+ \partial^2_x -a\partial_x-m^2)\phi,
\end{aligned}
\end{equation}
where we have used the fact that the change of metric is
\begin{align}
h_{\mu \nu}=
\begin{pmatrix}
2ax&  0&\\
0&  0& 
\end{pmatrix}. 
\end{align}
Then assuming that the solutions of $\phi$ have the form of plane waves $e^{-i\omega t+ikx}$ as modes with different $k$ or $\omega$ would not interact with each other. Substituting the plane waves to \eqref{scalar eom}, we find the spectrum 
\begin{equation}
\omega^2=k^2+iak+m^2  
\end{equation}
which modified the original real spectrum with an imaginary term $iak$. Thus indirectly we have shown that at least one of $k$ and $\omega$ becomes complex after a change of frame in this case.
}

If this speculation could be true, then it explains why a topologically trivial state could become topologically  nontrivial observed by a non-inertial observer: the nontrivial topological structure is hidden in the complex momentum plane, which could only be detected by certain non-inertial observers. However, to show this further evidence is still required.

\end{itemize}

\noindent{\bf The holographic calculation.}

Now we show that we could get the same spectrum of the hydrodynamic modes as in \eqref{eigen} from the holographic calculation. Since the inertial frame solutions only depend on $k_{z}$, which means $ h_{\mu \nu} $ propagates along the $z$ direction, we would first need to rotate the system so that the wave vector stays in an arbitrary direction $\overrightarrow{k} =(k_{x} ,k_{y} ,k_{z} )$.
$ h_{\mu \nu} $ transforms as $ h'_{\mu \nu } =\frac{\partial x^{\alpha } }{\partial x'^{\mu } } \frac{\partial x^{\beta  } }{\partial x'^{\nu } }h_{\alpha \beta } $ under the rotation. 
The transformation matrix $\frac{\partial x^{\mu } }{\partial x'^{\nu } } $ needed here is a three dimensional rotation that first rotates $\theta$ about the $y$ axis and then rotate $\phi $ about the $z$ axis, which is
\begin{align}
R=
\begin{pmatrix}
1&  0&  0&  0& 0\\
0&  \cos\phi\cos\theta&  -\sin\theta&  \cos\phi\sin\theta& 0\\
0&  \sin\phi\cos\theta&  \cos\phi&  \sin\phi\sin\theta& \\
0&  -\sin\theta&  0&  \cos\theta& 0\\
0&  0&  0&  0& 1
\end{pmatrix}   
\end{align}
The two rotation angle $\theta$ and $\phi$ are related to the new wave vector $\overrightarrow{k} =(k_{x} ,k_{y} ,k_{z} )$ by 
\begin{align}
\cos\theta=\frac{k_z}{\sqrt{k_x^2+k_y^2+k_z^2}},\\
\sin\theta=\frac{\sqrt{k_x^2+k_y^2}}{\sqrt{k_x^2+k_y^2+k_z^2}},\\
\cos\phi=\frac{k_x}{\sqrt{k_x^2+k_y^2}},\\
\sin\phi=\frac{k_y}{\sqrt{k_x^2+k_y^2}}.
\end{align}
After the rotation, we follow the procedure as in the Lorentz case. First we perform the coordinate transformation \eqref{CT2}.The rotated metric perturbation transforms again according to 
\begin{align}
h'_{\mu \nu } (x)=h_{\mu \nu } (x)-h_{\alpha \nu } (x)\partial_\mu \xi ^{\alpha }- h_{\mu\beta } (x)\partial_\nu \xi ^{\beta }-\xi ^{\lambda }\partial_\lambda h_{\mu \nu } (x).  
\end{align} 
After the transformations above we get the new coordinate space expression of $h'_{\mu \nu } (t,x,y,z,u)$. Then we transform it to the momentum space by Fourier transformation. Under the Fourier transformation, expressions like $x^j h_{\mu \nu }$ and $\partial_{x_j}h_{\mu \nu }$ transform as 
\begin{align}
t\to -i\partial_\omega ,x_j\to i\partial_{k_j}, \partial t\to-i\omega ,\partial_{x_j}\to ik_j.
\end{align} 
We only perform Fourier transformations in the $t, x, y, z$ directions, and the radial coordinate $u$ stays unchanged. Note that there is some singularity for this rotation at $k_x=k_y=0$ as $\phi$ becomes undetermined. To solve this problem, we assume that the rotation angles are fixed in the calculation when we take partial derivatives of $k^j$. For this to be correct as the final result we need to work in the limit that $k_y$ and $k_z$ are both zero. The reason is that in this limit, the two angles are indeed constant not depending on $k$'s and this limit is also the most  interesting limit for us to get the spectrum as the nontrivial physics lies in the $k_x$ direction. But we need to keep all $k_y$ and $k_z$ terms in the expressions until all derivatives have been correctly taken, otherwise some terms would get lost if $k_y$ and $k_z$ had been taken to be zero at the beginning.  

Using the rules above, we obtain the metric perturbation in the momentum space $h_{\mu\nu}(\omega,k_x,k_y,k_z,u)$. Then by setting $u=0$, the boundary value $h_{\mu\nu}(\omega,k_x,k_y,k_z,0)$ is obtained and we can calculate the poles of the integration constants following our prescription. This amounts to calculating the determinant of the matrix $M$.
In the lowest order of $\omega, k^j, m$ and $b$ and considering only the $k_x$ direction, the poles are determined by the following equation
\begin{align}\label{detM}
-\frac{1}{3}(b^2-3\omega^2)(k_x^2+m^2-3\omega^2)=0,
\end{align}
where note that $k_y$ and $k_z$ are set to 0. Due to the conformal invariance of the boundary theory, we have $v_s=\frac{1}{\sqrt{3}}$. Thus \eqref{detM} is equivalent to
\begin{align}
(b^2 v_s^2-\omega^2)(k_x^2 v_s^2+m^2 v_s^2-\omega^2)=0
\end{align}
The result here is identical to \eqref{eigen} (setting $k_y=k_z=0$) of the field theory calculation in section \ref{sec 2}.
This confirms that in the holographic system in the same non-inertial reference frame, we could obtain the desired result of spectrum that matches the field theory computation.

%
\section{Topological hydrodynamic modes with a $U(1)$ charge }\label{sec 4}

Up until now, we have studied the situation of zero $U(1)$ charge, i.e. we only considered the simplest hydrodynamic system with only a conserved energy momentum tensor. One may wonder if more complex topological structure could exist for hydrodynamic systems with more conserved currents. In this section, we generalize this discussion to the case with an extra $U(1)$ conserved current. We will try to weakly break all symmetries to see if interesting topological structure exists. In this work we will not discuss how these symmetry breaking effects could be produced and only focus on the possible resulting spectrum.

We consider an extra conservation current $J^\mu$ in addition to the energy momentum tensor in the following. We choose the hydrodynamic variables to be the fluctuation in the energy density $\delta \epsilon=T^{00}$, charge density $\delta n=J^0$ and momentum density $\pi_i=T^{0i}$. Without adding conservation breaking terms, after linearizing the five conservation equations around the equilibrium state, the effective Hamiltonian, as was defined by $i \partial \psi = H \psi$, ($\psi=(\delta n, \delta \epsilon,\pi_x,\pi_y,\pi_z)^T$), gives
\begin{align}
\begin{pmatrix}
0&  0&  \frac{\bar{n}}{\bar{w}} k_x&  \frac{\bar{n}}{\bar{w}} k_y& \frac{\bar{n}}{\bar{w}} k_z\\
0&  0&  k_x&  k_y& k_z\\
\beta_2 k_x&  \beta_1 k_x&  0&  0& 0\\
\beta_2 k_y&  \beta_1 k_y&  0&  0& 0\\
\beta_2 k_z&  \beta_1 k_z&  0&  0& 0
\end{pmatrix} 
\end{align}
at first order in the small $k,~\omega$ expansion, where $\beta_1=(\frac{\partial p}{\partial \epsilon})_n $, $\beta_2=(\frac{\partial p}{\partial n})_\epsilon $, $\bar{n}$ and $\bar{w}$ are the charge and enthalpy density in the equilibrium.

As a first step to deform the spectrum, we would directly modify the effective Hamiltonian to open a gap in the spectrum.
We require that the eigenvalues of the effective Hamiltonian should be real, which constrains the possible modifications.
In section \ref{sec 2}, we added two $m$ terms, i.e. \eqref{t1} and \eqref{t2} to gap the system. Here, with an extra conserved current, there exist three independent different ways of breaking the conservative equations. In each way, we have a mass parameter and we denote the three mass parameters to be $m_1$, $m_2$ and $m_3$. To have new band crossings in the spectrum, we still need the same $b$ terms as in section \ref{sec 2}.  In the following we will write out the extra non-conservation terms and the effective Hamiltonians for the three cases respectively.

\begin{itemize}
\item I. As the first case, we will directly generalize the system in section \ref{sec 2} to the system with an extra conserved $U(1)$ whose conservation is not broken, i.e. the current $J^{\mu }$ is conserved while $T^{\mu \nu}$ is not conserved with the same symmetry breaking terms as in section. \ref{sec 2}. We have
\begin{align}
&\partial_{\mu} \delta J^{\mu }  =  0 \label{J}\\
&\partial_{\mu} \delta T^{\mu t}  =  m_1 \delta T^{t x} \label{T1}\\
&\partial_{\mu} \delta T^{\mu x}  =  -m_1 v_{s}^{2} \delta T^{t t} \label{T2}\\
&\partial_{\mu} \delta T^{\mu y}  =  b \delta T^{t z}\label{T3}\\
&\partial_{\mu} \delta T^{\mu z}  =  -b\delta T^{t y}.\label{T4}
\end{align} 
The symmetry breaking of this kind could be produced by the small reference frame transformation \eqref{CT0}.
This is because those five equations are exactly the covariant conservation equation for $\delta J^{\mu }$ and $\delta T^{\mu \nu}$. The conservation equations for $\delta T^{\mu \nu}$ stay the same as in section \ref{sec 2}, while the covariant conservation equation for $\delta J^{\mu }$ is
\begin{align}
\partial_\mu \delta J^\mu + \Gamma^{\mu}_{\mu \lambda} \delta J^\lambda = 0.
\end{align}
As the transformation \eqref{CT0} has a traceless metric variation $h=0$, we have
\begin{align}
\Gamma^{\mu}_{\mu \lambda} = \frac{1}{2} \partial_\alpha h = 0 .
\end{align} Thus the reference frame transformations do not affect the conservation of $J^{\mu }$, i.e. \eqref{J} still holds after the transformation \eqref{CT0}.
The effective Hamiltonian is then modified to be 
\begin{align}\label{H1}
\begin{pmatrix}
               0&                     0& \frac{\bar{n}}{\bar{w}} k_x&  \frac{\bar{n}}{\bar{w}} k_y& \frac{\bar{n}}{\bar{w}} k_z\\
               0&                     0&                 k_x + i m_1&                          k_y&                         k_z\\
     \beta_2 k_x& \beta_1 (k_x - i m_1)&                           0&                            0&                           0\\
     \beta_2 k_y&           \beta_1 k_y&                           0&                            0&                         i b\\
     \beta_2 k_z&           \beta_1 k_z&                           0&                         -i b&                           0
\end{pmatrix} 
\end{align}

The mass effect lies in the interaction between the energy and the $x$ direction momentum. The system in this case is a direct generalization of section \ref{sec 2} with an extra conserved $U(1)$ current. Similar to that case, the mass term between the energy and the $y$ (or $z$) direction momentum, which we denote $m_y$ (or $m_z$) would gap the system, and in this sense the resulting spectrum is a symmetry protected gapless topological state. As we will show later topologically non-trivial modes similar to \eqref{sp} with one extra zero mode would be found in this system.

\item II.
{The second way to break the conservation and find interesting gapless topological modes is as follows}
\begin{equation}
\begin{aligned}
&\partial_{\mu} \delta J^{\mu }  =  m_2 \frac{\bar{n}}{\bar{w}} T^{t x} \\
&\partial_{\mu} \delta T^{\mu t}  =  0 \\
&\partial_{\mu} \delta T^{\mu x}  =  -m_2 \beta_2 \delta J^{t } \\
&{\partial_{\mu} \delta T^{\mu y}  =  b \delta T^{t z}}\\
&{\partial_{\mu} \delta T^{\mu z}  =  -b\delta T^{t y}}.
\end{aligned} 
\end{equation}
{The corresponding effective Hamiltonian is
\begin{align}
\begin{pmatrix}
                    0&                     0& \frac{\bar{n}}{\bar{w}} (k_x + i m_2)&  \frac{\bar{n}}{\bar{w}} k_y& \frac{\bar{n}}{\bar{w}} k_z\\
                    0&                     0&                                   k_x&                          k_y&                         k_z\\
\beta_2 (k_x - i m_2)&           \beta_1 k_x&                                     0&                            0&                           0\\
          \beta_2 k_y&           \beta_1 k_y&                                     0&                            0&                         i b\\
          \beta_2 k_z&           \beta_1 k_z&                                     0&                         -i b&                           0
\end{pmatrix} .
\end{align}
} The mass effect lies in the interaction between the $U(1)$ current and the $x$ direction momentum. 

\item III.
{The third way to break the conservation and find interesting gapless topological modes is as follows}
\begin{equation}
\begin{aligned}
&\partial_{\mu} \delta J^{\mu }  =  -m_3 \delta T^{t t} \\
&\partial_{\mu} \delta T^{\mu t}  =  m_3 \delta J^{t } \\
&\partial_{\mu} \delta T^{\mu x}  =  0 \\
&{\partial_{\mu} \delta T^{\mu y}  =  b \delta T^{t z}}\\
&{\partial_{\mu} \delta T^{\mu z}  =  -b\delta T^{t y}}
\end{aligned}.
\end{equation}
{The corresponding effective Hamiltonian is}
\begin{align}
\begin{pmatrix}
                    0&                 i m_3& \frac{\bar{n}}{\bar{w}} k_x &  \frac{\bar{n}}{\bar{w}} k_y& \frac{\bar{n}}{\bar{w}} k_z\\
               -i m_3&                     0&                                   k_x&                          k_y&                         k_z\\
          \beta_2 k_x&           \beta_1 k_x&                                     0&                            0&                           0\\
          \beta_2 k_y&           \beta_1 k_y&                                     0&                            0&                         i b\\
          \beta_2 k_z&           \beta_1 k_z&                                     0&                         -i b&                           0
\end{pmatrix} 
\end{align} The mass effect lies in the interaction between the $U(1)$ current and the energy. 
\end{itemize}

Note that we cannot find a reference frame transformation to explain the second or the third way of symmetry breaking just as we did in the first way. The mechanism to produce these symmetry breaking effects is left for future work.

Each of these three ways of modification would produce a similar effect in the spectrum. This modified spectrum is similar to \eqref{sp} but has one more flat band as we have five variables now. Besides that additional flat band, it is qualitatively the same as what we discussed in the case of zero $U(1)$ charge in section \ref{sec 2}. Since the three ways above have qualitative the same effect, we consider the first one here as an example to study the topological property of its spectrum.

Note that in this system, there are four nodes when $m_1<b$ and the same as in section \ref{sec 2}, these nodes are topologically nontrivial protected by the mirror symmetry $M$: $y \to -y$, $z \to -z$, because the nodes would disappear if there are $m_y$ or $m_z$ terms in $y$ and $z$ directions. We could calculate the topological invariants of this system in two ways as introduced in seciton \ref{subsec 2.3}, which are in fact equivalent in essence. 

\begin{figure}[htbp]
    \centering
    \includegraphics[scale=0.5]{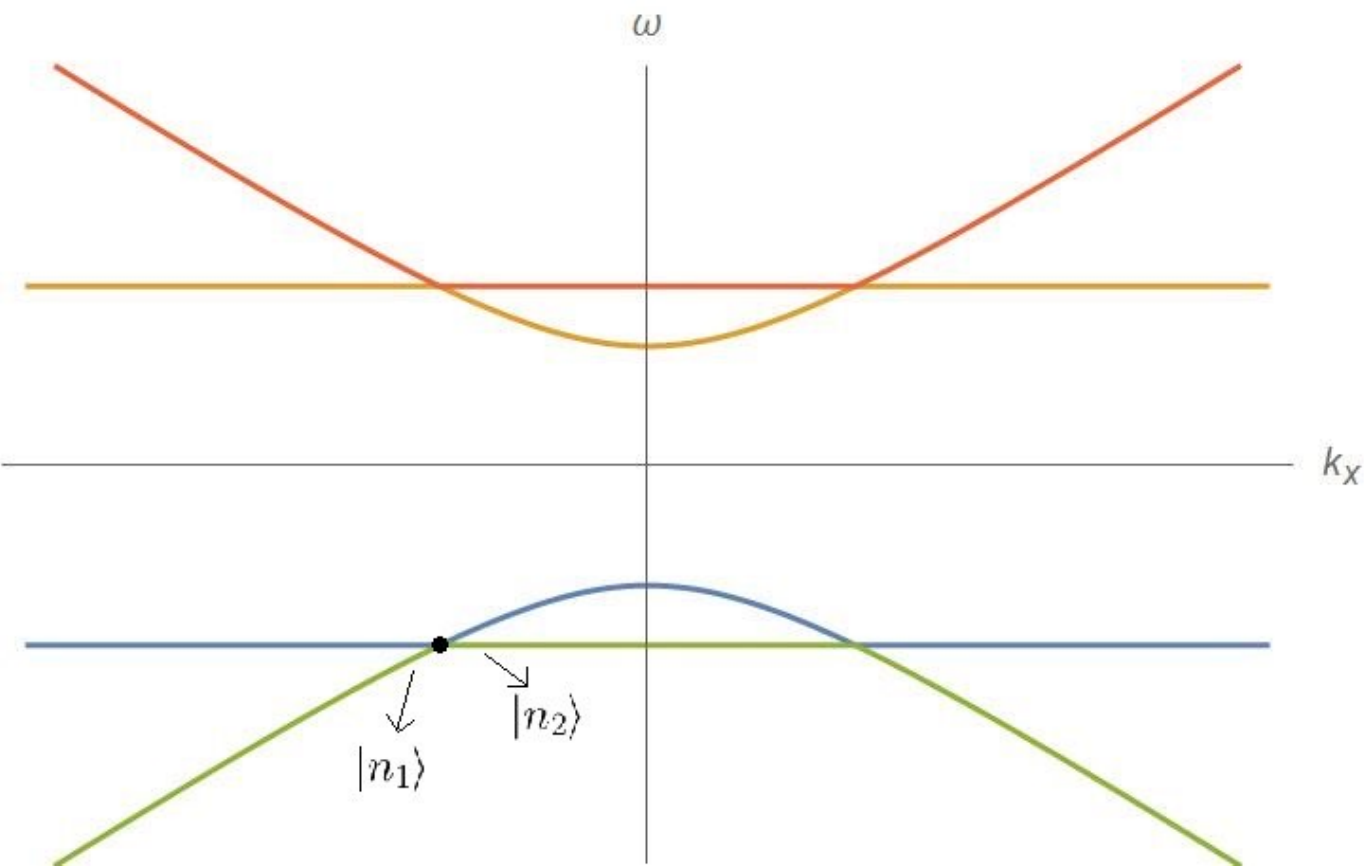}
    \caption{The spectrum \eqref{u1 spectrum}. }
    \label{u1 m1}
\end{figure}

As the topological state is a symmetry protected one, the topological invariants should be calculated at high symmetric points $k_y=0$, $k_z=0$ in the momentum space.
The spectrum at these high symmetric points is
\begin{align}\label{u1 spectrum}
\omega_1=0, \;\; \omega_{2,3}=\pm b,  \;\; \omega_{4,5}=\pm \sqrt{k_x^2( \beta_1+\frac{\bar{n}}{\bar{w}} \beta_2 )+ m_1^2 \beta_1}
\end{align}
as shown in figure \ref{u1 m1}.
Again we consider the left lower node as an example as the four nodes are equivalent. We denote the left limit state to be $|n_1 \rangle$ and right limit state to be $|n_2 \rangle$.
$|n_1 \rangle$ and $|n_2 \rangle$ are nothing but two eigenvectors that correspond to the two lower bands of the Hamiltonian \eqref{H1}.

{\footnotesize
\begin{align}
\left | n_1  \right \rangle &= \sqrt{\frac{m_1^2 \beta_1^2+k_x^2(\beta_1+(\bar{n}/\bar{w}) \beta_2)}{(1+(\bar{n}/\bar{w})^2)k_x^2+m_1^2}}(-\frac{n k_x}{\sqrt{k_x^2 (\beta_1+(\bar{n}/\bar{w}) \beta_2)+m_1^2 \beta_1}},\frac{k_x + i m_1}{\sqrt{k_x^2 (\beta_1+(\bar{n}/\bar{w}) \beta_2)+m_1^2 \beta_1}},1,0,0)\\
\left | n_2  \right \rangle &= \frac{1}{\sqrt{2}}(0,0,0,i,1)
\end{align}}

{The first way to calculate the topological invariant is from $\xi$ which is defined as the difference between the number of occupied bands that have eigenvalue $+1$ of the symmetry $M$ at the left and right limits of the node. Since $|n_1 \rangle$ has eigenvalue $+1$ under the mirror symmetry and $|n_2 \rangle$ has eigenvalue $-1$. We have $\xi=+1$ which is a non-trivial value, implying the node is a topologically nontrivial one protected by the symmetry $M$.}

The second way to define the topological invariant following section \ref{sec 2} is the normalized inner product of the  $|n_1 \rangle$ and $|n_2 \rangle$. The two eigenvectors are obviously orthonormal to each other. Thus the Berry phase of the node is undetermined, indicating that the node is a topologically nontrivial one. Both of the topological invariants defined above indicate that the nodes in this charged case are topologically non-trivial protected by the $M$ symmetry.

It is also easy to see that the change in the relative magnitude of $m_1$ and $b$ would again gap the nodes and cause a topological phase transition between trivial gapped and topologically nontrivial gapless states.

Besides considering the three cases above separately, we could also turn on the three mass terms altogether. When this mixture of the three modifications is considered, there would be one more interesting spectrum, which is qualitatively different from \eqref{sp}.

\begin{figure}[htbp]
    \centering
    \subfigure[]{
    \begin{minipage}[b]{.3\linewidth}
    \centering
    \includegraphics[scale=0.3]{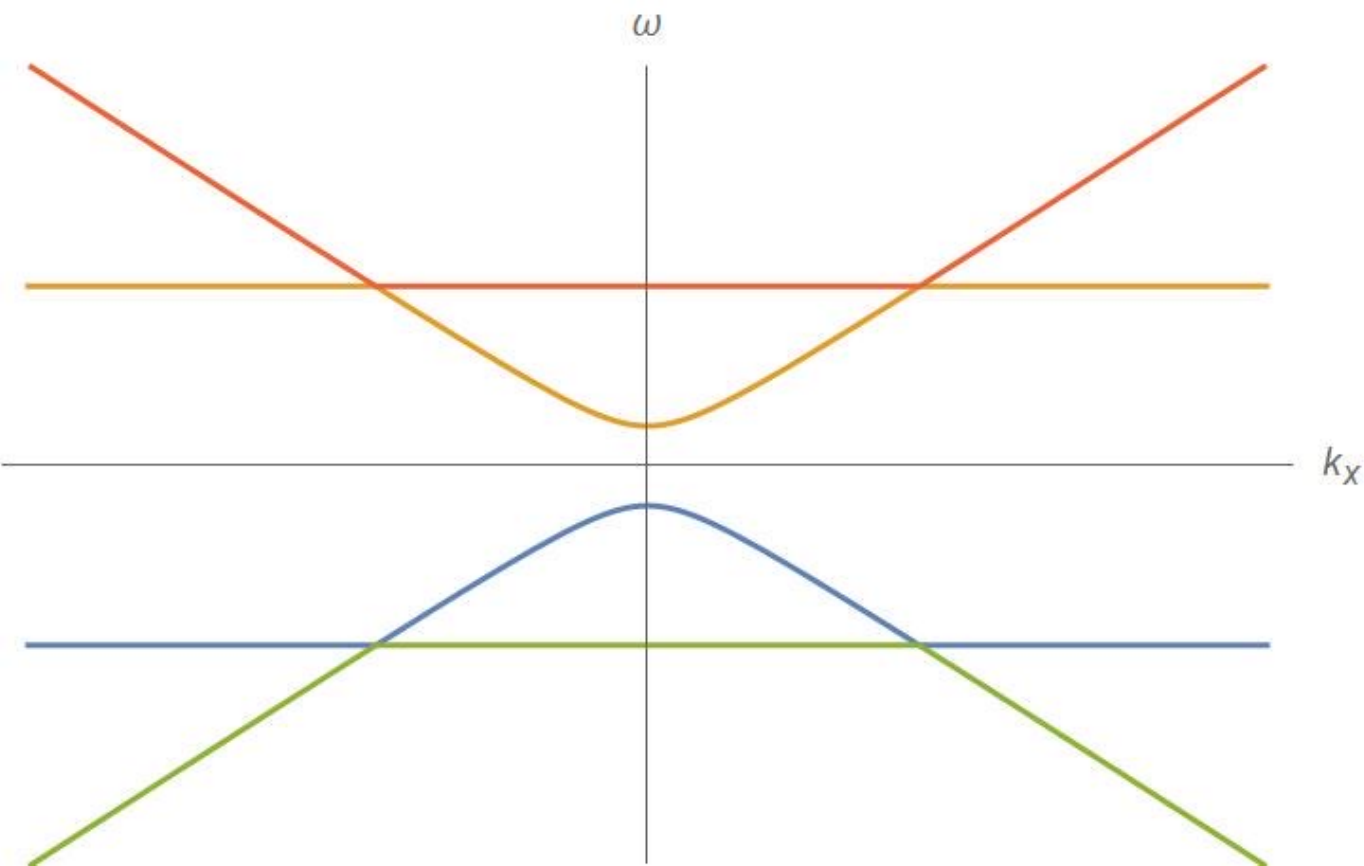}
    \end{minipage}
    }
    \subfigure[]{
    \begin{minipage}[b]{.3\linewidth}
    \centering
    \includegraphics[scale=0.3]{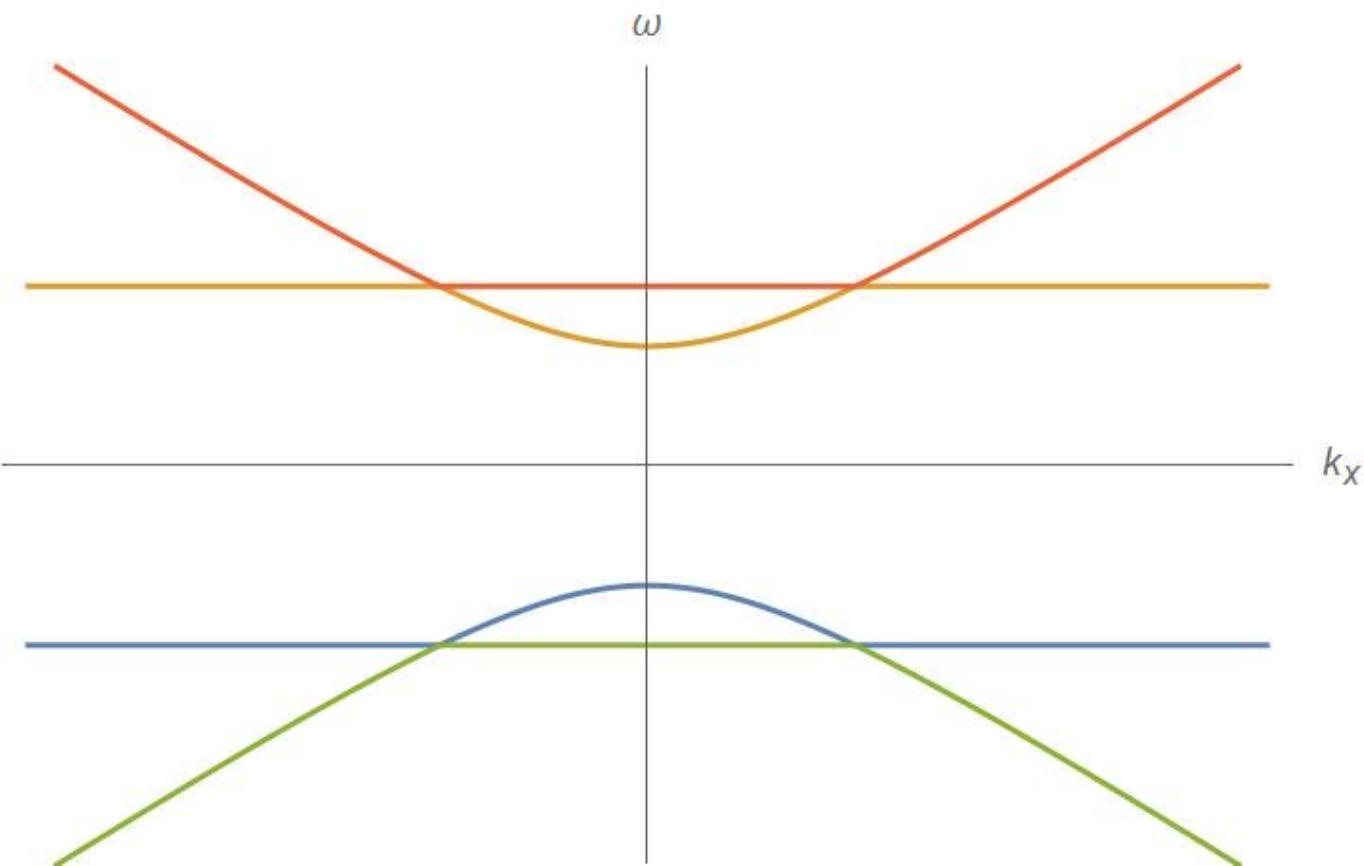}
    \end{minipage}
    }
    \caption{The combination effect of $m_1$ and $m_2$ terms in the spectrum of the modified hydrodynamics with a conserved U(1) current. (a) $m_1 \ne 0$ and $m_2=0$; (b) $m_2$ is turned on to a non-zero value while $m_1$ takes the same value as $m_1$ in (a). The gap becomes larger in (b) indicating that the effective mass parameter becomes larger.}
    \label{u1 m1m2}
\end{figure}

\begin{figure}[htbp]
    \centering
    \subfigure[]{
    \begin{minipage}[b]{.3\linewidth}
    \centering
    \includegraphics[scale=0.25]{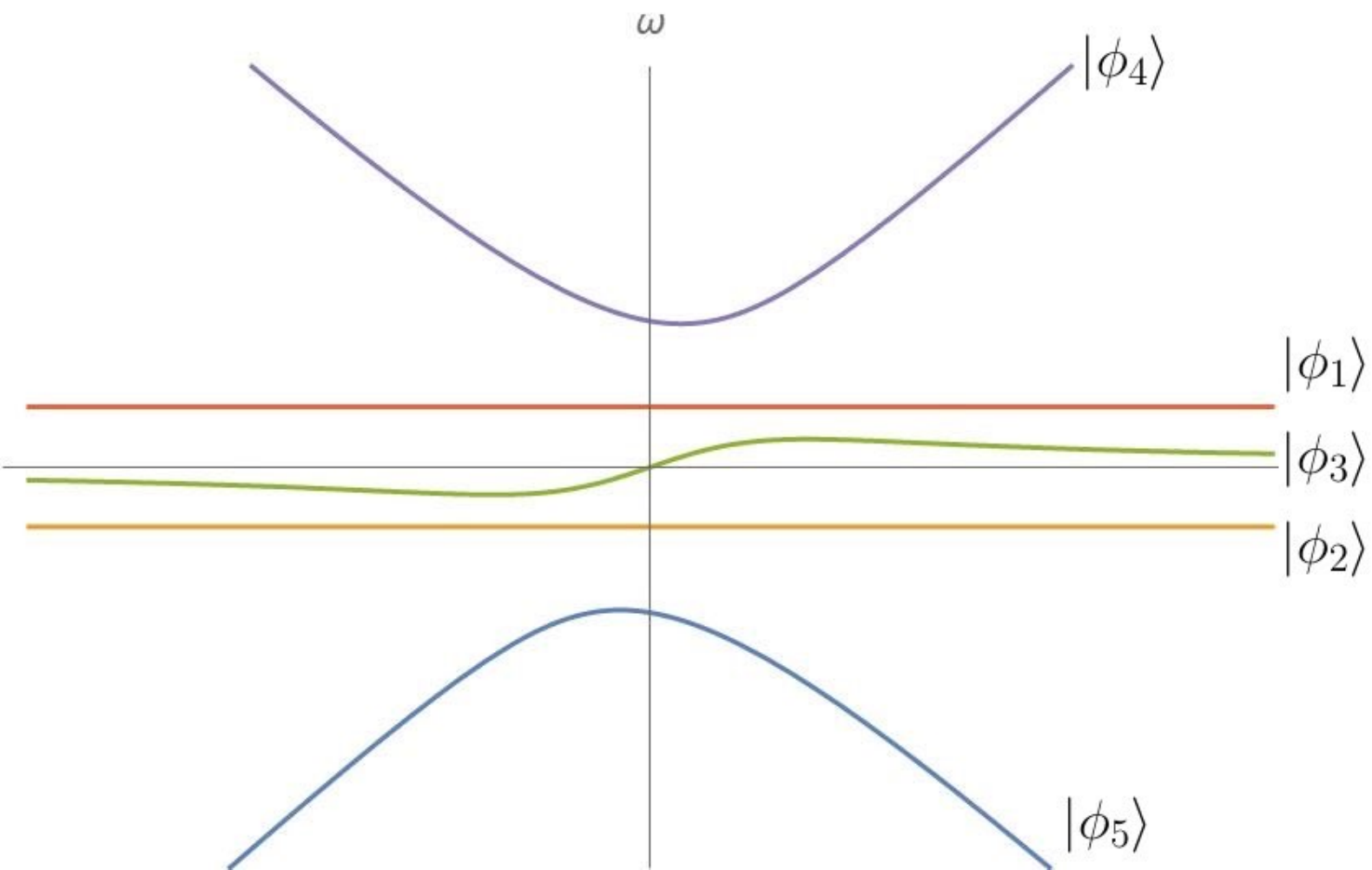}
    \end{minipage}
    }
    \subfigure[]{
    \begin{minipage}[b]{.3\linewidth}
    \centering
    \includegraphics[scale=0.25]{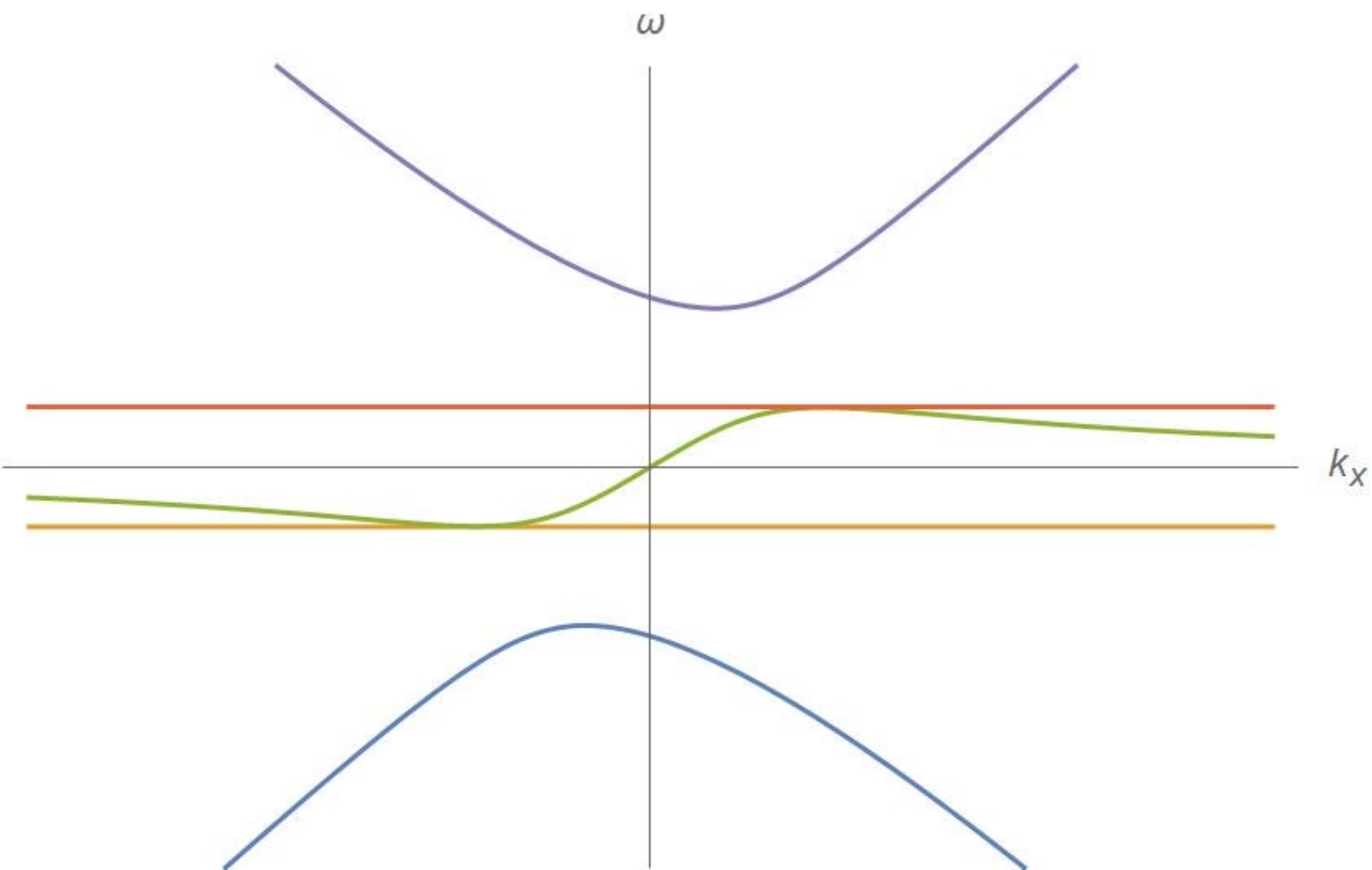}
    \end{minipage}
    }
    \subfigure[]{
    \begin{minipage}[b]{.3\linewidth}
    \centering
    \includegraphics[scale=0.25]{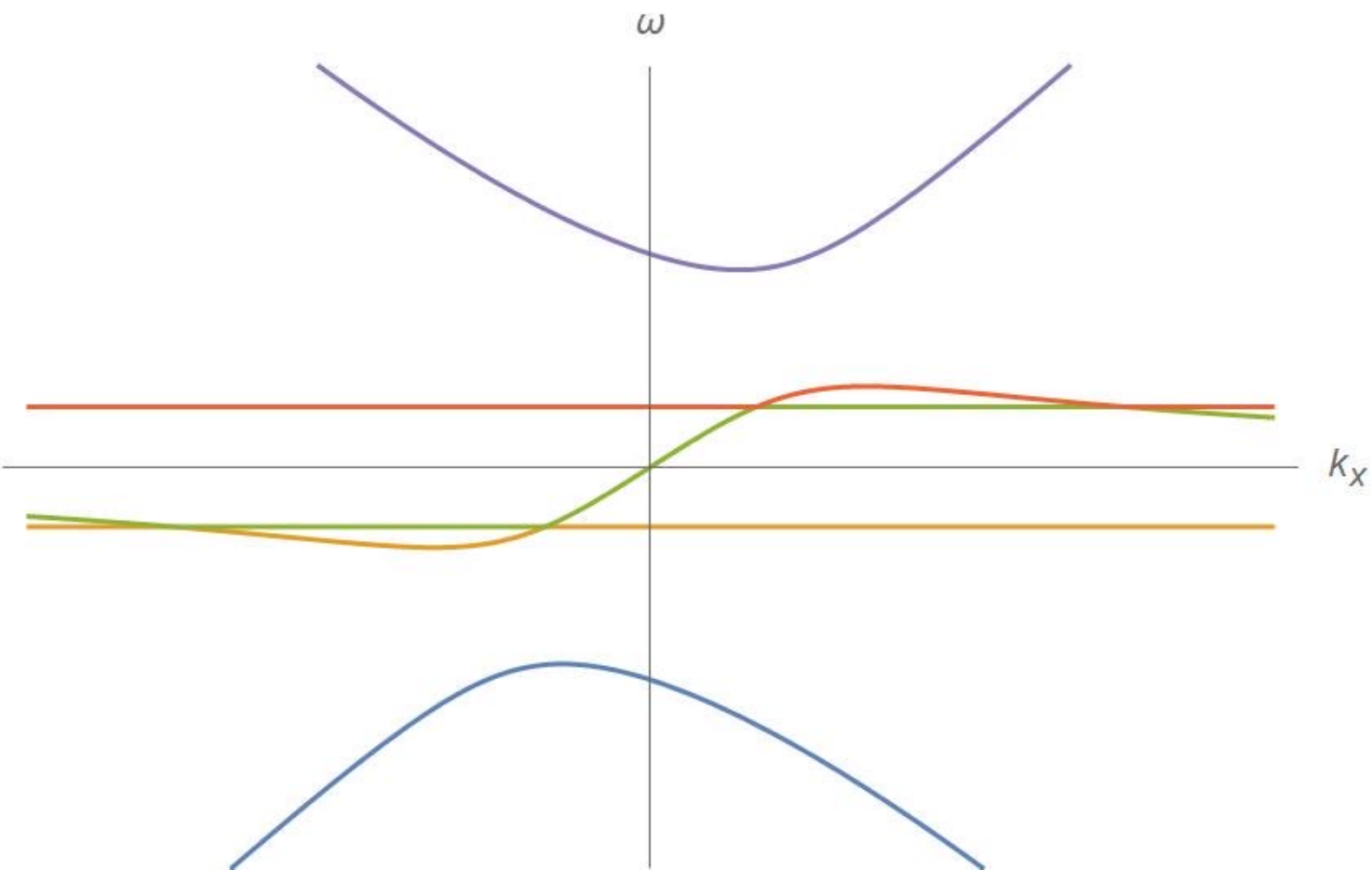}
    \end{minipage}
    }
    
    \subfigure[]{
    \begin{minipage}[b]{.3\linewidth}
    \centering
    \includegraphics[scale=0.25]{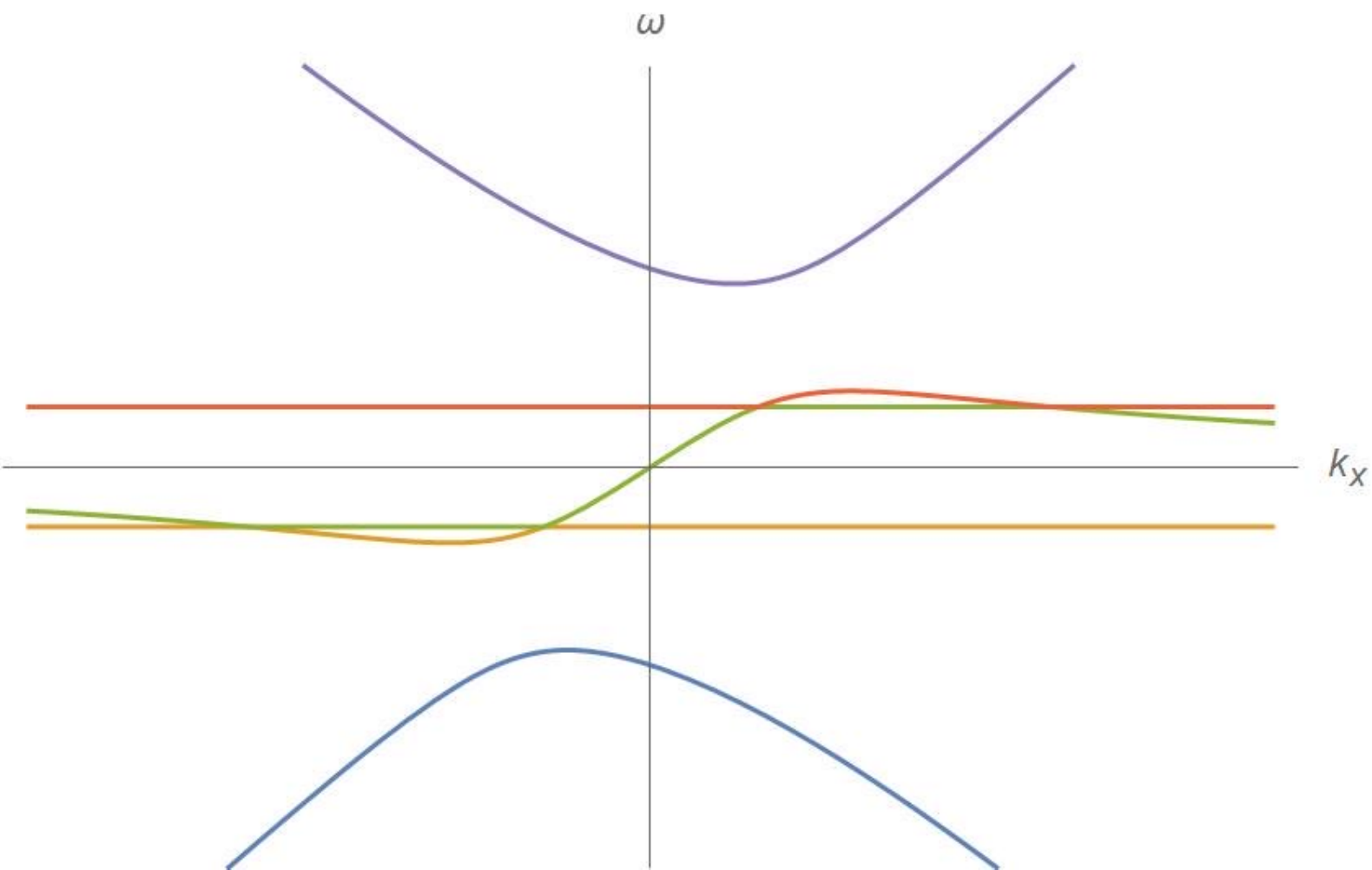}
    \end{minipage}
    }
    \subfigure[]{
    \begin{minipage}[b]{.3\linewidth}
    \centering
    \includegraphics[scale=0.25]{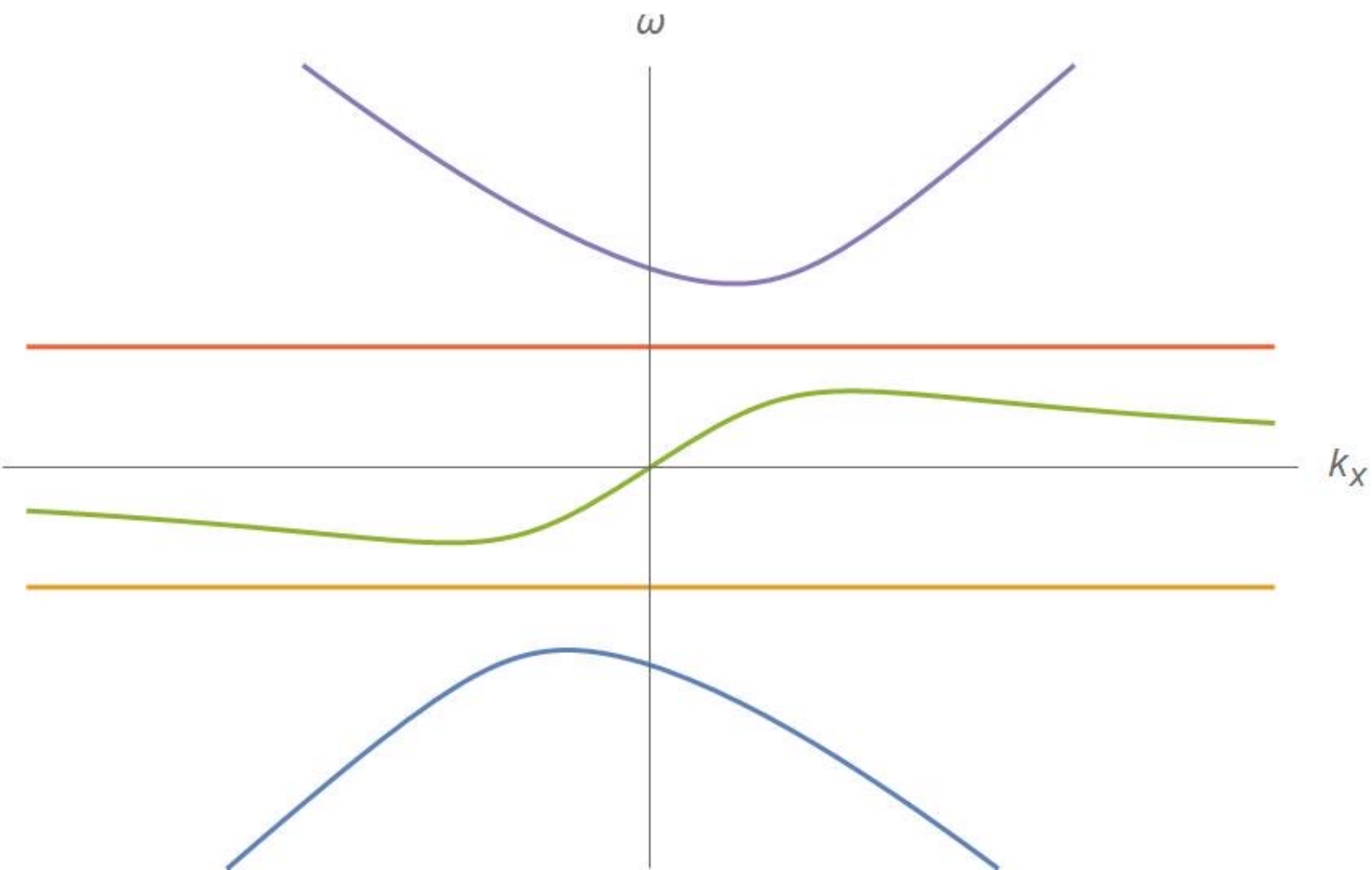}
    \end{minipage}
    }
    \subfigure[]{
    \begin{minipage}[b]{.3\linewidth}
    \centering
    \includegraphics[scale=0.25]{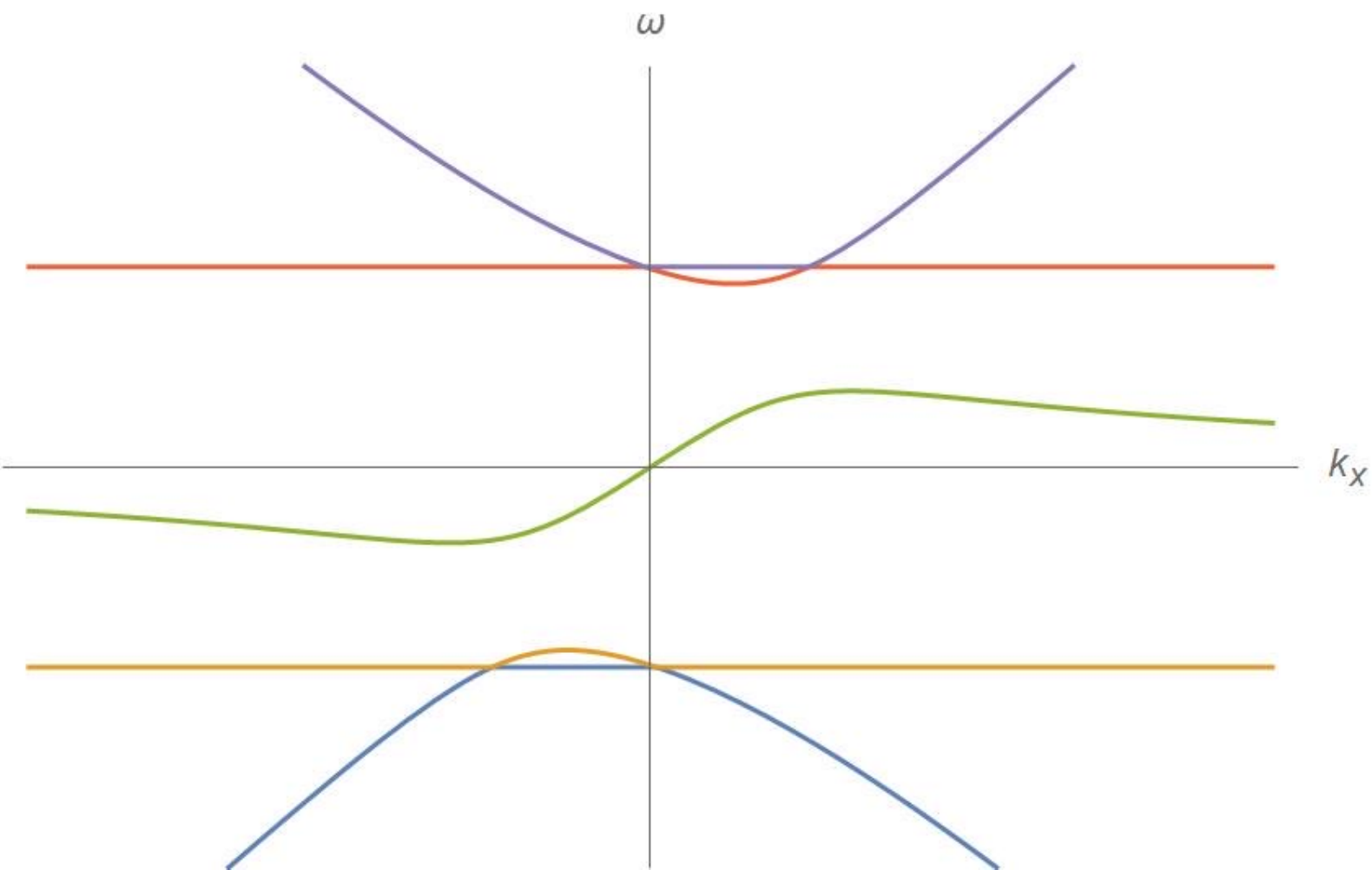}
    \end{minipage}
    }
    \caption{The combination effect of $m_1$ and $m_3$ terms in the spectrum of the modified hydrodynamics with a finite U(1) charge. $m_1$ and $m_3$ together determine the extent to which the three curved bands deviate from the axial symmetry of $\omega$ axis while $b$ determines the distance between the two flat bands. 
    From (a) to (c): The value of $m_3$ changes from small to large with fixed $m_1$ and $b$. The change of $m_3$ distorts the three curved bands away from axial symmetry and result in topological phase transition. As $m_3$ increases from 0, at first there is no crossing ((a), excluding the $\omega=k=0$ point), then after a critical point (b) with two band crossing nodes, the number of nodes becomes four in (c). From (d) to (f): The value of $b$ changes from small to large with fixed $m_1$ and $m_3$. The band crossing behaviour depends on the value of $b$. As b increases from 0, at first there are four band crossing nodes (d), then no crossing (e) and finally four crossing nodes again (f).}
    \label{u1 m1m3}
\end{figure}

Let us consider all the possible combinations of the $m_1$, $m_2$ and $m_3$ terms. The effective Hamiltonian becomes
\begin{align}
\begin{pmatrix}
                    0&                 i m_3&  \frac{\bar{n}}{\bar{w}} (k_x + i m_2)&  \frac{\bar{n}}{\bar{w}} k_y& \frac{\bar{n}}{\bar{w}} k_z\\
               -i m_3&                     0&                            k_x + i m_1&                          k_y&                         k_z\\
\beta_2 (k_x - i m_2)& \beta_1 (k_x - i m_1)&                                      0&                            0&                           0\\
          \beta_2 k_y&           \beta_1 k_y&                                      0&                            0&                         i b\\
          \beta_2 k_z&           \beta_1 k_z&                                      0&                         -i b&                           0
\end{pmatrix} 
\end{align}
The parameters $m_1$, $m_2$ and $m_3$ could be turned on separately or jointly. The effect of the combination of $m_1$, $m_2$ and the combination of $m_1$, $m_3$ are shown in figure. \ref{u1 m1m2} and figure. \ref{u1 m1m3} respectively. As we can see in the figures, the effect of combining $m_1$ and $m_2$ is the same as applying any one of the three mass terms and only increases the effective mass parameter when an extra mass parameter is turned on. However, the combination of $m_1$ and $m_3$ as well as the combination of $m_2$ and $m_3$ would lead to a new spectrum as shown in figure \ref{u1 m1m3}.

To study the topological property of this new spectrum, we implement again the Berry phase computation. This system is again topologically protected by the mirror symmetry $M$. Thus the Berry phase is calculated at the high symmetry points $k_y=k_z=0$. There are five bands, among which two are flat bands and three are curved bands. Each of the five bands represents an eigen-state of the effective Hamiltonian. The five eigen-states at general $k_x$ have the following form
\begin{equation}
\begin{aligned}
\left | \phi_1  \right \rangle &= (0,0,0,-i,1)\\
\left | \phi_2  \right \rangle &= (0,0,0,i,1)\\
\left | \phi_3  \right \rangle &= (f_1(k_x),f_2(k_x),1,0,0)\\
\left | \phi_4  \right \rangle &= (g_1(k_x),g_2(k_x),1,0,0)\\
\left | \phi_5  \right \rangle &= (h_1(k_x),h_2(k_x),1,0,0)
\end{aligned},
\end{equation}
where $\left | \phi_1  \right \rangle$ and $\left | \phi_2  \right \rangle$ represent the upper and lower flat band respectively, while the other three states correspond to the three curved bands, see figure \ref{u1 m1m3}(a).
$f_1$, $f_2$, $g_1$, $g_2$, $h_1$, $h_2$ are functions of $k_x$, whose explicit forms are not important here.
The Berry phase of all the nodes in figure \ref{u1 m1m3} are undetermined since both $\left | \phi_1  \right \rangle$ and $\left | \phi_2  \right \rangle$ are orthonormal to $\left | \phi_3  \right \rangle$, $\left | \phi_4  \right \rangle$ and $\left | \phi_5  \right \rangle$.
This concludes that all the nodes are topologically non-trivial protected by the reflection symmetry $M$, including the four nodes of (c) in figure \ref{u1 m1m3} as well as the four nodes of (f) in figure \ref{u1 m1m3}.

In summary, there are more mass terms one can add to the effective Hamiltonian to get similar topologically nontrivial gapless modes. The mass terms when added together may change the shape of the spectrum and lead to more complicated structure in the topological phase diagrams. However, the calculation of the topological invariants shows that all the band crossing nodes still possess the same topological property as in section \ref{sec 2}, i.e. they are topologically nontrivial nodes protected by the reflection symmetry.

%
\section{Conclusions and discussions }\label{sec 5}

In this paper, based on the work in  \cite{Liu:2020ksx,Liu:2020abb} we have calculated the holographic hydrodynamics modes with the same topological spectrum as in \cite{Liu:2020ksx,Liu:2020abb} in a non-inertial frame version of holography and generalized the topological hydrodynamic system to the case with an extra $U(1)$ symmetry. The previous works had shown that weakly breaking the conservation of the energy momentum tensor could produce a topological gapless hydrodynamic system in relativistic hydrodynamics. One way to obtain this weak breaking of energy momentum conservation is to observe the system by an accelerating observer, i.e. changing from an inertial frame to a specific non-inertial one.

With the aim to construct a holographic realization of the topological hydrodynamic modes, as a first step, \cite{Liu:2020abb} confirmed that the Ward identities for the energy momentum tensor could be reproduced from the holographic system of the same non-inertial reference frame. As a further step, in this paper, we computed the spectrum of the modes explicitly in this holographic system and found that the holographic computation matched the previous results directly calculated in hydrodynamics. To perform this computation, we have developed a prescription to compute the poles of two point Green's function in a different reference frame without the need to calculate the perturbations in the new metric background of the new frame. We also generalized the study of topological modes in relativistic hydrodynamics to the case with one extra $U(1)$ current and found that more complicated topological phase diagrams could exist when we consider more possibilities of the mass terms.

However, there are still some open questions left, e.g. the physical meaning of the symmetry breaking in the U(1) case may not be interpreted by a frame transformation. Thus the possible mechanism to realize such symmetry breaking terms in this case still requires further investigation. Also we have discussed the mechanism of why topologically trivial states could become topologically nontrivial in a non-inertial frame. Though physically in a non-inertial frame, new forces would appear in the equation of motion, which could produce topologically nontrivial states, it is still an open question to understand this from a more mathematical aspect, i.e. on-shell states would still be on-shell states after a reference frame transformation, thus where is this new topologically nontrivial state in the original inertial frame? Our speculation is that this nontrivial topological structure was hidden in the complex plane of the original spectrum, which is supported by several pieces of indirect evidence, but further direct evidence is still required. Other open questions include if this property of topologically nontrivial states being observed in a non-inertial frame could be found also for other systems, e.g. fermions \cite{pan2022}, scalars coupled with other systems, vectors, etc., and if massive gravity or other symmetry breaking methods in holography could also produce similar topological hydrodynamic modes.

%
\section*{Acknowledgements }\label{sec 6}
We would like to thank Hyun-Sik Jeong, Xuan-Ting Ji, Yan Liu, Yuan-Tai Wang, Pei-Hung Yuan, Long Zhang for valuable discussions. This work was supported by the National Key R\&D Program of China (Grant No. 2018FYA0305800), Project 12035016 supported by National Natural Science Foundation of China, the Strategic Priority Research Program of Chinese Academy of Sciences, Grant No. XDB28000000.    


\begin{thebibliography}{99}

\bibitem{lu2014topological}
L. Lu, J. Joannopoulos and M. Soljačić, 
{\em Topological photonics,}
\href{https://doi.org/10.1038/nphoton.2014.248}{Nature Photon 8, 821–829 (2014)}. 

\bibitem{ozawa2019topological}
T. Ozawa, H. M. Price, A. Amo, N. Goldman et al.,
{\em Topological photonics,}
\doi{10.1103/RevModPhys.91.015006}{Rev. Mod. Phys. 91, 015006}.

\bibitem{zhang2018topological}
X. Zhang, M. Xiao, Y. Cheng, M.-H. Lu and J. Christensen,  
{\em Topological sound,}  
\doi{10.1038/s42005-018-0094-4}{Nature Communications 1, 97 (2018)}. 

\bibitem{delplace2017topological}
P. Delplace, J. B. Marston, and A. Venaille,  
{\em Topological origin of equatorial waves,}  
\doi{10.1126/science.aan8819}{Science 358, 1075-1077 (2017)}. 
[\arXiv{1702.07583}{cond-mat.mes-hall}].

\bibitem{souslov2019topological}
A.~Souslov, K.~Dasbiswas, M.~Fruchart, S.~Vaikuntanathan and V.~Vitelli,
{\em Topological waves in fluids with odd viscosity,}
\doi{10.1103/PhysRevLett.122.128001}{Phys. Rev. Lett. \textbf{122}, no.12, 128001 (2019)}
[\arXiv{1802.09649} {cond-mat.soft}].

\bibitem{green2020topological}
R. Green, J. Armas, J. de Boer and L. Giomi,
{\em Topological waves in passive and active fluids on curved surfaces: a unified picture,}
[\arXiv{2011.12271}{cond-mat.soft}].

\bibitem{Kovtun:2012rj} 
P.~Kovtun,
{\em Lectures on hydrodynamic fluctuations in relativistic theories,}
\doi{10.1088/1751-8113/45/47/473001}{J.\ Phys.\ A {\bf 45}, 473001 (2012)}
[\arXiv{1205.5040}{hep-th}].

\bibitem{Liu:2020ksx}
Y.~Liu and Y.~W.~Sun,
{\em Topological modes in relativistic hydrodynamics,}
\doi{10.1103/PhysRevD.103.044044}{Phys. Rev. D \textbf{103}, no.4, 044044 (2021)}
[\arXiv{2004.13380} {hep-th}].

\bibitem{Liu:2020abb}
Y.~Liu and Y.~W.~Sun,
{\em Topological hydrodynamic modes and holography,}
\doi{10.1103/PhysRevD.105.086017}{Phys. Rev. D \textbf{105}, no.8, 086017 (2022)}
[\arXiv{2005.02850} {hep-th}].

\bibitem{Policastro:2002se}
G.~Policastro, D.~T.~Son and A.~O.~Starinets,
{\em From AdS / CFT correspondence to hydrodynamics,}
\doi{10.1088/1126-6708/2002/09/043}{JHEP \textbf{09}, 043 (2002)}
[\arXiv{hep-th/0205052} {hep-th}].


\bibitem{Policastro:2002tn}
G.~Policastro, D.~T.~Son and A.~O.~Starinets,
{\em From AdS / CFT correspondence to hydrodynamics. 2. Sound waves,}
\doi{10.1088/1126-6708/2002/12/054}{JHEP \textbf{12}, 054 (2002)}
[\arXiv{hep-th/0210220}{hep-th}].

\bibitem{Liu:2018djq}
Y.~Liu and Y.~Sun,
{\em Topological invariants for holographic semimetals,}
\doi{10.1007/JHEP10(2018)189}{JHEP \textbf{10}, 189 (2018)}
[\arXiv{1809.00513}{hep-th}].

\bibitem{Landsteiner:2019kxb}
K.~Landsteiner, Y.~Liu and Y.~Sun,
{\em Holographic Topological Semimetals,}
\doi{10.1007/s11433-019-1477-7}{Sci. China Phys. Mech. Astron. \textbf{63}, no.5, 250001 (2020)} 
[\arXiv{1911.07978}{hep-th}].

\bibitem{fang2016topological}
C. Fang, H. Weng, X. Dai and Z. Fang, 
{\em Topological nodal line semimetals}, 
\doi{10.1088/1674-1056/25/11/117106}{Chin. Phys. B 25,117106 (2016)} 
[\arXiv{1609.05414}{cond-mat.mes-hall}].

\bibitem{Liu:2018bye}
Y.~Liu and Y.~W.~Sun,
{\em Topological nodal line semimetals in holography,}
\doi{10.1007/JHEP12(2018)072}{JHEP \textbf{12} (2018), 072}
[\arXiv{1801.09357}{hep-th}].

\bibitem{Zaanen:2015oix}
J.~Zaanen, Y.~W.~Sun, Y.~Liu and K.~Schalm,
{\em Holographic Duality in Condensed Matter Physics,}
\doi{10.1017/CBO9781139942492}{Cambridge University Press (2015)}

\bibitem{Son:2007vk}
D.~T.~Son and A.~O.~Starinets,
{\em Viscosity, Black Holes, and Quantum Field Theory,}
\doi{10.1146/annurev.nucl.57.090506.123120}{Ann. Rev. Nucl. Part. Sci. \textbf{57}, 95-118 (2007)}
[\arXiv{0704.0240}{hep-th}].

\bibitem{Hubeny:2011hd}
V.~E.~Hubeny, S.~Minwalla and M.~Rangamani,
{\em The fluid/gravity correspondence,}
[\arXiv{1107.5780}{hep-th}].

\bibitem{Gubser:1998bc}
S.~S.~Gubser, I.~R.~Klebanov and A.~M.~Polyakov,
{\em Gauge theory correlators from noncritical string theory,}
\doi{10.1016/S0370-2693(98)00377-3}{Phys. Lett. B \textbf{428}, 105-114 (1998)}
[\arXiv{hep-th/9802109} {hep-th}].

\bibitem{Witten:1998qj}
E.~Witten,
{\em Anti-de Sitter space and holography,}
\doi{10.4310/ATMP.1998.v2.n2.a2}{Adv. Theor. Math. Phys. \textbf{2}, 253-291 (1998)}
[\arXiv{hep-th/9802150} {hep-th}].

\bibitem{moller1952theory}
C.~Moller,
{\em The Theory of Relativity,}
{Clarendon Press, Oxford (1952)}.

\bibitem{pan2022}
W.~B.~Pan and Y.~W.~Sun,
{\em work in progress}.



\end{thebibliography}
\end{document}